\documentclass{aa}
\usepackage{graphicx}
\usepackage{threeparttable}
\def \la{\mathrel{\mathchoice   {\vcenter{\offinterlineskip\halign{\hfil
$\displaystyle##$\hfil\cr<\cr\sim\cr}}}
{\vcenter{\offinterlineskip\halign{\hfil$\textstyle##$\hfil\cr
<\cr\sim\cr}}}
{\vcenter{\offinterlineskip\halign{\hfil$\scriptstyle##$\hfil\cr
<\cr\sim\cr}}}
{\vcenter{\offinterlineskip\halign{\hfil$\scriptscriptstyle##$\hfil\cr
<\cr\sim\cr}}}}}
\def \ga{\mathrel{\mathchoice   {\vcenter{\offinterlineskip\halign{\hfil
$\displaystyle##$\hfil\cr>\cr\sim\cr}}}
{\vcenter{\offinterlineskip\halign{\hfil$\textstyle##$\hfil\cr
>\cr\sim\cr}}}
{\vcenter{\offinterlineskip\halign{\hfil$\scriptstyle##$\hfil\cr
>\cr\sim\cr}}}
{\vcenter{\offinterlineskip\halign{\hfil$\scriptscriptstyle##$\hfil\cr
>\cr\sim\cr}}}}}
\def\vlsr {\hbox{${v_{\rm LSR}}$}}
\def\delv {\hbox{$\Delta v_{1/2}$}}

\begin{document}
\title{Dense gas in nearby galaxies}
\subtitle{XVI. The nuclear starburst environment in NGC\,4945 {\thanks {Based on observations
with the Swedish/ESO Submillimeter Telescope (SEST) at the European Southern
Observatory (ESO), La Silla, Chile}} }
\author{M.~Wang\inst{1,2} \and C.~Henkel\inst{1} \and Y.-N. Chin\inst{3}
\and J.~B.~Whiteoak\inst{4} \and M.~Hunt~Cunningham\inst{5} \and R.~Mauersberger\inst{6}
\and D.~Muders\inst{1}}
\offprints{M.~Wang (mwang@pmo.ac.cn)}
\institute{
  Max-Planck-Institut f{\"u}r Radioastronomie,
  Auf dem H{\"u}gel 69, D-53121 Bonn, Germany
\and
  Purple Mountain Observatory, Chinese Academy of Sciences,
  210008 Nanjing, China
\and
        Department of Physics, Tamkang University,
        251-37 Tamsui, Taipei County, Taiwan
\and
  Australia Telescope National Facility, CSIRO Radiophysics Labs.,
  P.O. Box 76, Epping, NSW 2121, Australia
\and
  School of Physics, UNSW, 2052 Sydney, Australia
\and
  IRAM, Avenida Divina Pastora 7, Local 20, E-18012 Granada, Spain }
\date{Received date ; accepted date}
\abstract{ A multi-line millimeter-wave study of the nearby
starburst galaxy NGC\,4945 has been carried out using the
Swedish-ESO Submillimeter Telescope (SEST). The study covers the
frequency range from 82\,GHz to 354\,GHz and includes 80
transitions of 19 molecules. 1.3~mm continuum data of the nuclear
source are also presented. An analysis of CO and 1.3~mm continuum
fluxes indicates that the conversion factor between H$_2$ column
density and CO $J$=1--0 integrated intensity is smaller than in
the galactic disk by factors of 5--10. A large number of molecular
species indicate the presence of a prominent high density
interstellar gas component characterized by $n_{\rm
H_2}\sim$10$^5$\,cm$^{-3}$. Some spectra show Gaussian profiles.
Others exhibit two main velocity components, one at
$\sim$450\,km\,s$^{-1}$, the other at $\sim$710\,km\,s$^{-1}$.
While the gas in the former component has a higher linewidth, the
latter component arises from gas that is more highly excited as is
indicated by HCN, HCO$^+$ and CN spectra. Abundances of molecular
species are calculated and compared with abundances observed
toward the starburst galaxies NGC\,253 and M\,82 and galactic
sources. Apparent is an `overabundance' of HNC in the nuclear
environment of NGC\,4945. While the HNC/HCN $J$=1--0 line
intensity ratio is $\sim$0.5, the HNC/HCN abundance ratio is
$\sim$1. From a comparison of K$_{\rm a}$=0 and 1 HNCO line
intensities, an upper limit to the background radiation of 30\,K
is derived. While HCN is subthermally excited ($T_{\rm
ex}$$\sim$8\,K), CN is even less excited ($T_{\rm
ex}$$\sim$3--4\,K), indicating that it arises from a less dense
gas component and that its $N$=2--1 line can be optically thin
even though its $N$=1--0 emission is moderately optically thick.
Overall, fractional abundances of NGC\,4945 suggest that the
starburst has reached a stage of evolution that is intermediate
between those observed in NGC\,253 and M\,82. Carbon, nitrogen,
oxygen and sulfur isotope ratios are also determined. Within the
limits of uncertainty, carbon and oxygen isotope ratios appear to
be the same in the nuclear regions of NGC\,4945 and NGC\,253. High
$^{18}$O/$^{17}$O, low $^{16}$O/$^{18}$O and $^{14}$N/$^{15}$N and
perhaps also low $^{32}$S/$^{34}$S ratios (6.4$\pm$0.3,
195$\pm$45, 105$\pm$25 and 13.5$\pm$2.5 in NGC\,4945,
respectively) appear to be characteristic properties of a
starburst environment in which massive stars have had sufficient
time to affect the isotopic composition of the surrounding
interstellar medium.  }

\maketitle

\section{Introduction}

NGC\,4945, a nearby edge-on ($i$=78$^{\circ}$) spiral starburst
galaxy at a systemic velocity of $V_{\rm sys}$ $\sim$
560\,km\,s$^{-1}$ (e.g. Ott et al. 2001), is one of the three
brightest IRAS point sources ($S_{100\mu \rm m}$ = 686\,Jy;
IRAS\,1989) beyond the Magellanic clouds and a goldmine for
molecular cloud research. NGC\,4945 contains a highly obscured
`compton thick' (e.g. Maiolino et al. 1999) Seyfert 2 nucleus and
is classified as SB(s)cd or SAB(s)cd. Being a member of the
Centaurus group, its distance was estimated to lie between
3.9\,Mpc (de Vaucouleurs 1964; Bergman et al. 1992) and 8.1\,Mpc
(Baan 1985). The value of 6.7\,Mpc is most often used (e.g. Dahlem
et al. 1993; Henkel et al. 1994a, hereafter H94; Ott et al. 2001),
but 4\,Mpc (20$''$ correspond to 390\,pc) is a more realistic
estimation (Bergman et al. 1992) and will be used throughout the
article.

Numerous studies from the cm-waveband to hard X-rays have been
made towards this prominent southern galaxy. Associated with the
nuclear region is a strong radio continuum source with
$S_{1.4\,\rm GHz}$=\,4.8\,Jy (Whiteoak \& Bunton, 1985; Ott et al.
2001) and the first known megamaser (Dos Santos \& L{\'e}pine
1979;  Batchelor et al. 1982), detected in the 22\,GHz water vapor
line. Numerous molecular emission lines were detected, initially
at cm and later at mm wavelengths (Whiteoak \& Gardner 1975;
Whiteoak \& Wilson 1990; Henkel et al. 1990; H94; Curran et al.
2001, hereafter C01). An edge-on nuclear gas ring with a size of
$\sim$\, 30$''$ was discovered by Whiteoak et al. (1990) in the
$^{12}$C$^{16}$O $J$=1--0 line. A detailed analysis of
$^{12}$C$^{16}$O and $^{13}$C$^{16}$O profiles by Bergman et al.
(1992) and additional CO observations by Dahlem et al. (1993) and
Mauersberger et al. (1996a) further support the presence of this
ring. An HCN map indicates that low and high density molecular gas
coexists in the central 30$''$ (H94). On a much smaller scale, the
distribution of H$_2$O megamaser emission suggests the presence of
a circumnuclear disk with a radius of 20 milliarcseconds and a
binding mass of $\sim$\,10$^6$ M$_{\odot}$ (Greenhill et al.
1997).

A nuclear source with a column density of $N$(H) $\sim$
4$\times$10$^{24}$\,cm$^{-2}$ and possibly disk-like morphology is
indicated by X-ray data (Done et al. 2003). Strongly processed
ices were detected near 4.65$\mu$m in the more extended nuclear
environment (Spoon et al. 2003). A conical wind-blown cavity was
revealed by ground based spectra and images at optical and IR
wavelengths (Nakai 1989; Moorwood et al. 1996). With the NICMOS
Camera on-board of the Hubble Space Telescope (HST) a starburst
ring of size 5$''$ was found in Pa\,$\alpha$; the walls of a
conical cavity blown by supernova driven winds were detected in
H$_2$ (Marconi et al. 2000). X-ray observations reveal a complex
morphology including a strong variable nuclear source at
2--10\,keV and a conically shaped `plume' at 0.3--2\,keV (Schurch
et al. 2002).

To better understand the nuclear environment of NGC\,4945, we present observations of a large
number of molecular lines towards its central region and calculate relative abundances, isotope
ratios and H$_2$ densities.

\begin{table}
\begin{threeparttable}
\begin{scriptsize}
\caption[]{Observational parameters}
\begin{flushleft}
\begin{tabular}{ccccc}
\hline
$\nu$ & $\theta_{\rm b}^{\rm a)}$ & $\theta_{\rm b}^{\rm a)}$ & $T_{\rm sys}^{\rm b)}$ &
                                                             $\eta_{\rm b}^{\rm c)}$ \\
(GHz)             & (arcsec)      & (pc)            & (K)                            \\
\hline

                  &               &                 &                 &              \\
 82--99           &  63--53       & 1220--1030      &  150--200       & 0.78--0.73   \\
100--153          &  52--34       & 1000--660       &  300--400       & 0.73--0.61   \\
154--230          &  33--22       &  640--420       &  600--1000      & 0.61--0.46   \\
241--267          &  21--19       &  410--370       & 1500--2300      & 0.44--0.41   \\
330--354          &  19--15       &  370--290       & $\sim$3000      & 0.32--0.30   \\
                  &               &                 &                 &              \\

\hline
\end{tabular}
\begin{tablenotes}
\item[a)] Full width to half power (FWHP) beam widths. To establish linear scales, $D$=4\,Mpc
          was adopted.
\item[b)] System temperatures in units of main beam brightness temperature ($T_{\rm mb}$)
\item[c)] Beam efficiencies derived from measurements of Jupiter (L. Knee, priv. comm.)
\end{tablenotes}
\end{flushleft}
\end{scriptsize}
\end{threeparttable}
\end{table}

\section{Observations}

\subsection{Spectroscopic observations}

The observations were carried out in May 1994, September 1995,
March and June 1996, January and July 1997, January and July 1998,
September 2001 and February 2003 with the 15-m SEST telescope at
La Silla. 3\,mm and 2\,mm receivers as well as 3\,mm and 1.3\,mm
receivers were employed simultaneously. For the observations at
330--354\,GHz, another SIS receiver was used that was described in
detail by Mauersberger et al. (1996a). Two acousto-optical
backends (AOS) each covered a bandwidth of $\sim$\,1\,GHz. With a
total of 1600 and 1440 channels, the channel spacing was
0.68\,MHz.

All measurements were made in a dual beam switching mode
(switching frequency 6\,Hz) with a beam throw of 11$'$40$''$ in
azimuth. Since rapid beam switching was used in conjunction with
reference positions on both sides of the galaxy, baselines are of
good quality. Calibration was obtained with the chopper wheel
method (for internal consistency and comparisons with other
telescopes, see e.g. Sect.\,3.2 and H94 (their Sect.\,2) and Chin
et al. (1996, their Sect.\,2)). Observed frequency ranges, beam
widths, system temperatures and beam efficiencies are given in
Table~1. The latter, measured at 94, 115, 230 and 345\,GHz, were
interpolated and, if necessary, also extrapolated to convert
antenna temperatures ($T_{\rm A}^{*}$) to a main beam brightness
temperature ($T_{\rm mb}$) scale. The pointing accuracy, obtained
from measurements of the nearby continuum source Cen\,A, was
mostly better than 10$''$ (see Sect.\,3.2).

\begin{figure*}[ht]
\resizebox{\hsize}{!}{\includegraphics[angle=-90]{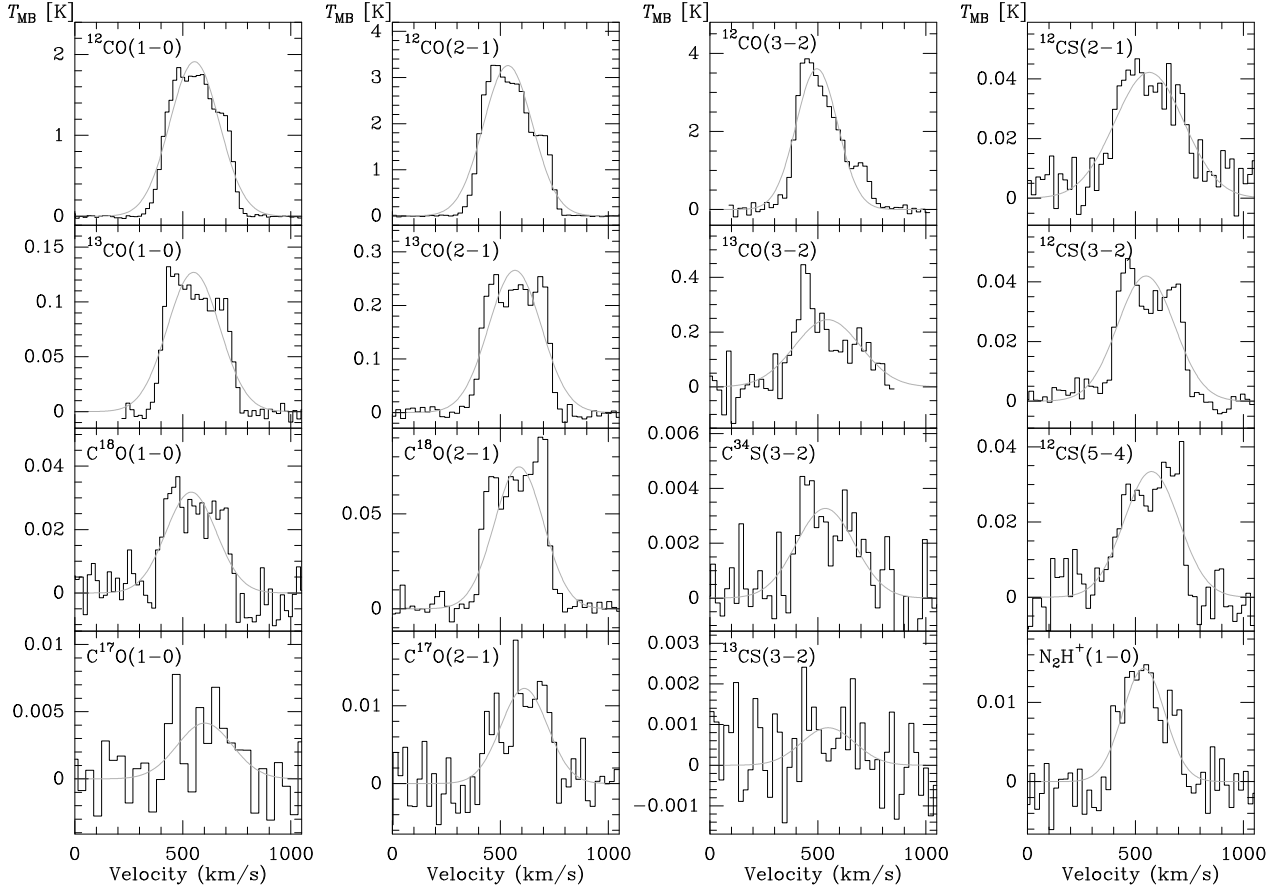}}
\caption[]{CO, CS and N$_2$H$^+$ spectra measured toward the
center of NGC\,4945. The velocity resolution for each spectrum is
$\sim\,$20\,km\,s$^{-1}$, the velocity scale is Local Standard of
Rest. Gaussian fits were made towards detected and tentatively
detected lines, and resulting parameters are listed in Table~2
(same in Figs.\,2--5).}
\end{figure*}

\begin{figure*}[ht]
\resizebox{\hsize}{!}{\includegraphics[angle=-90]{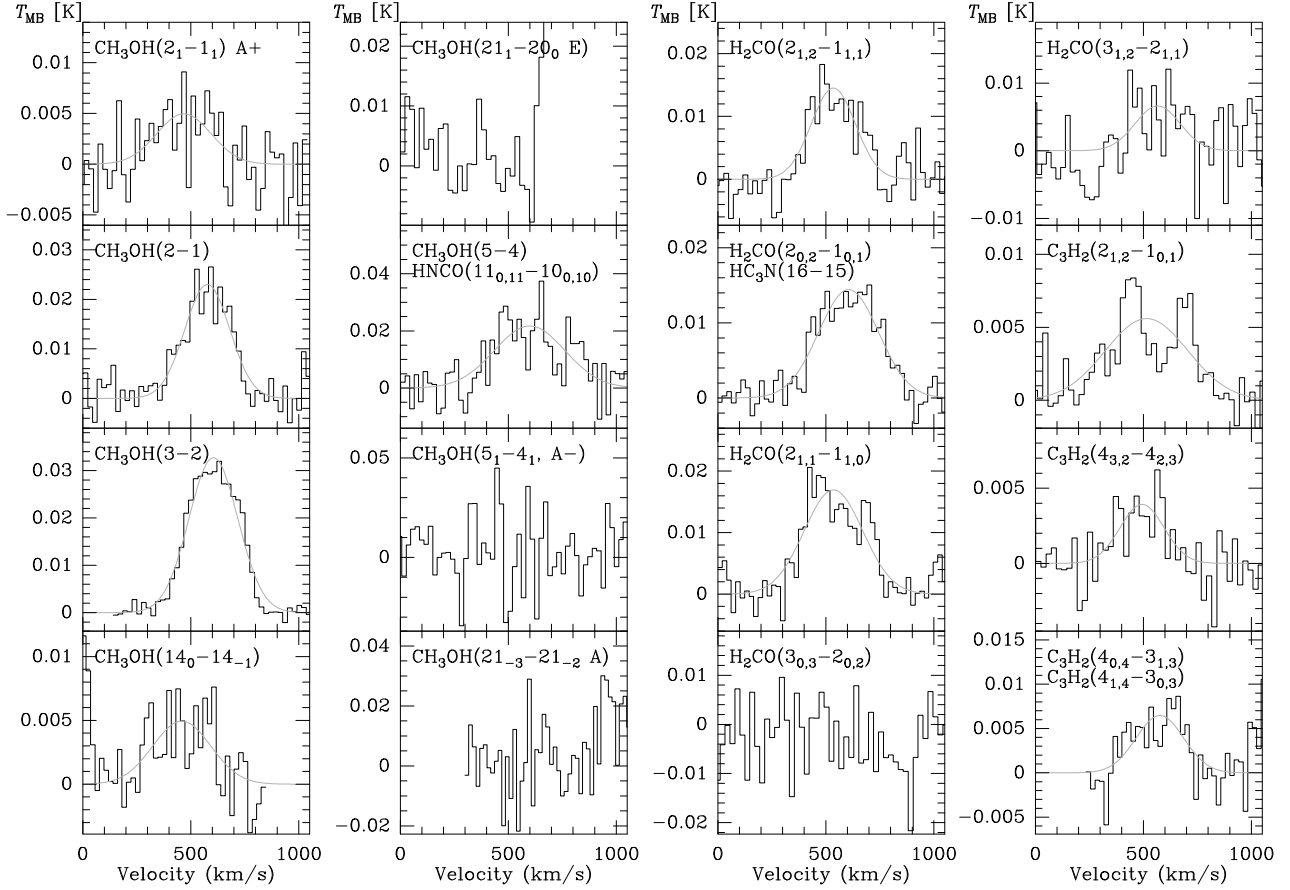}}
\caption[]{CH$_3$OH, H$_2$CO and C$_3$H$_2$ spectra measured
toward the center of NGC\,4945. The $V$-axis is always defined
according to the frequency of the first transition given in the
upper left corner of each box; corresponding frequencies are
listed in Table~2.}
\end{figure*}

\begin{figure*}[ht]
\resizebox{\hsize}{!}{\includegraphics[angle=-90]{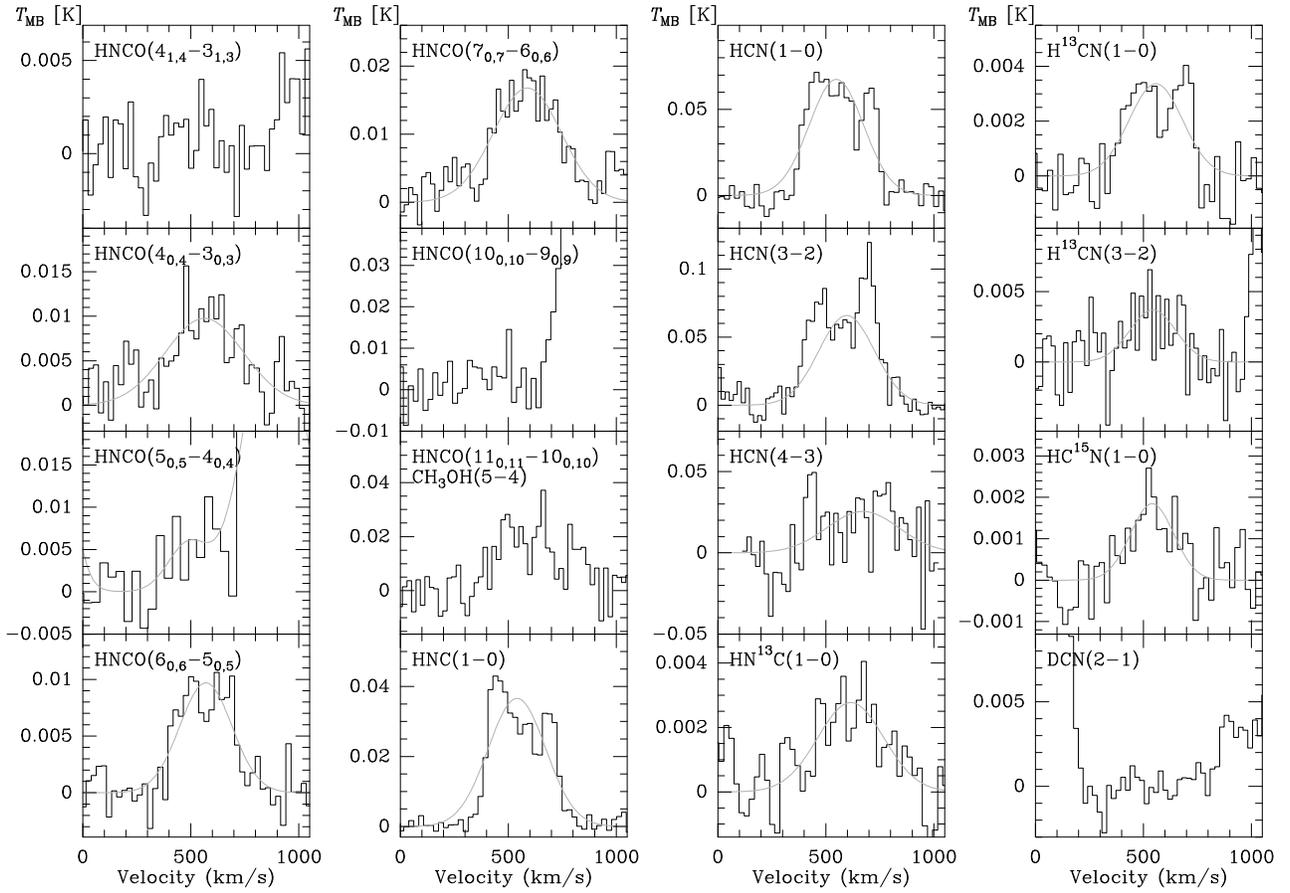}}
\caption[]{HNCO, HCN and HNC spectra measured toward the center of
NGC\,4945 (see also the captions to Figs.\,1 and 2). }
\end{figure*}

\begin{figure*}[ht]
\resizebox{\hsize}{!}{\includegraphics[angle=-90]{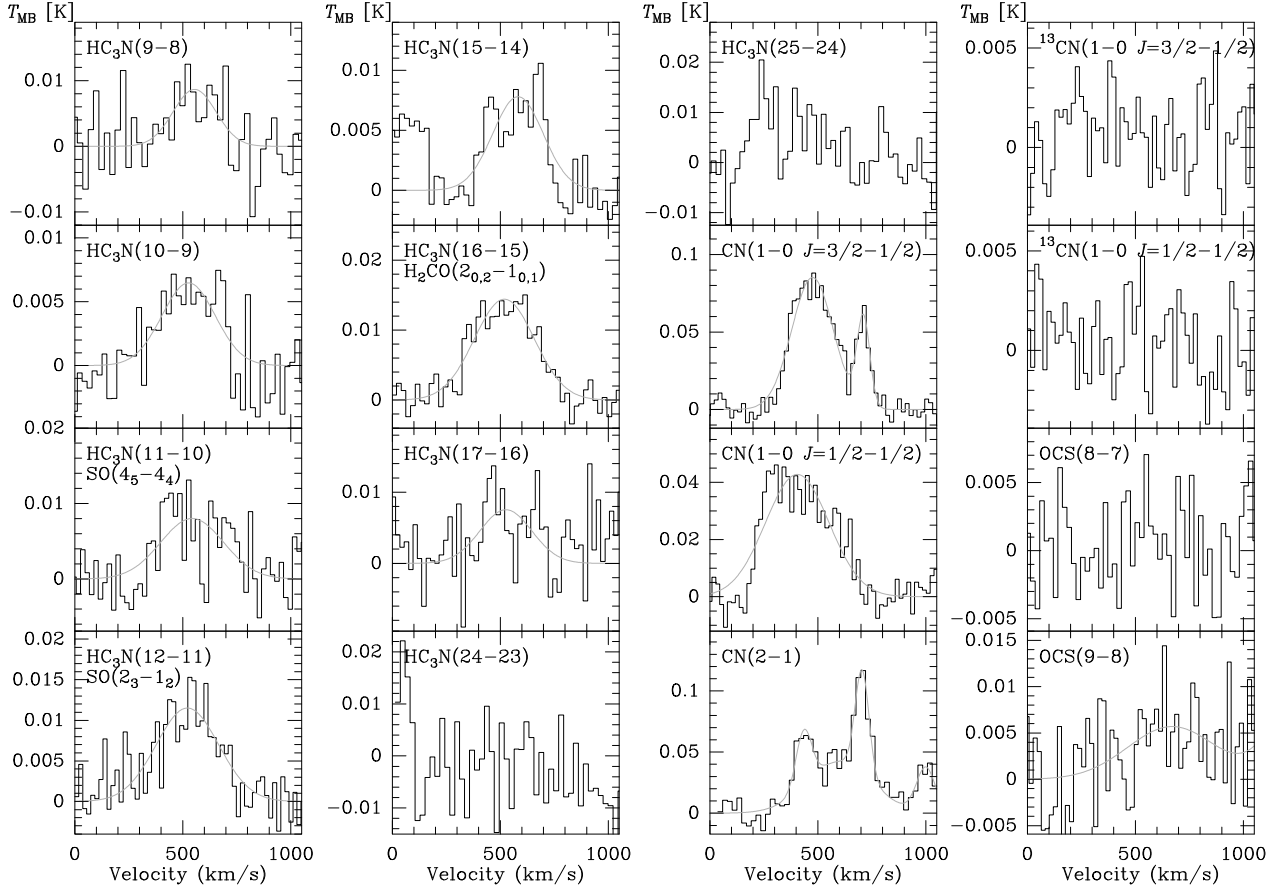}}
\caption[]{HC$_3$N, CN and OCS spectra measured toward the center
of NGC\,4945 (see also the captions to Figs.\,1 and 2). }
\end{figure*}

\begin{figure*}[ht]
\resizebox{\hsize}{!}{\includegraphics[angle=-90]{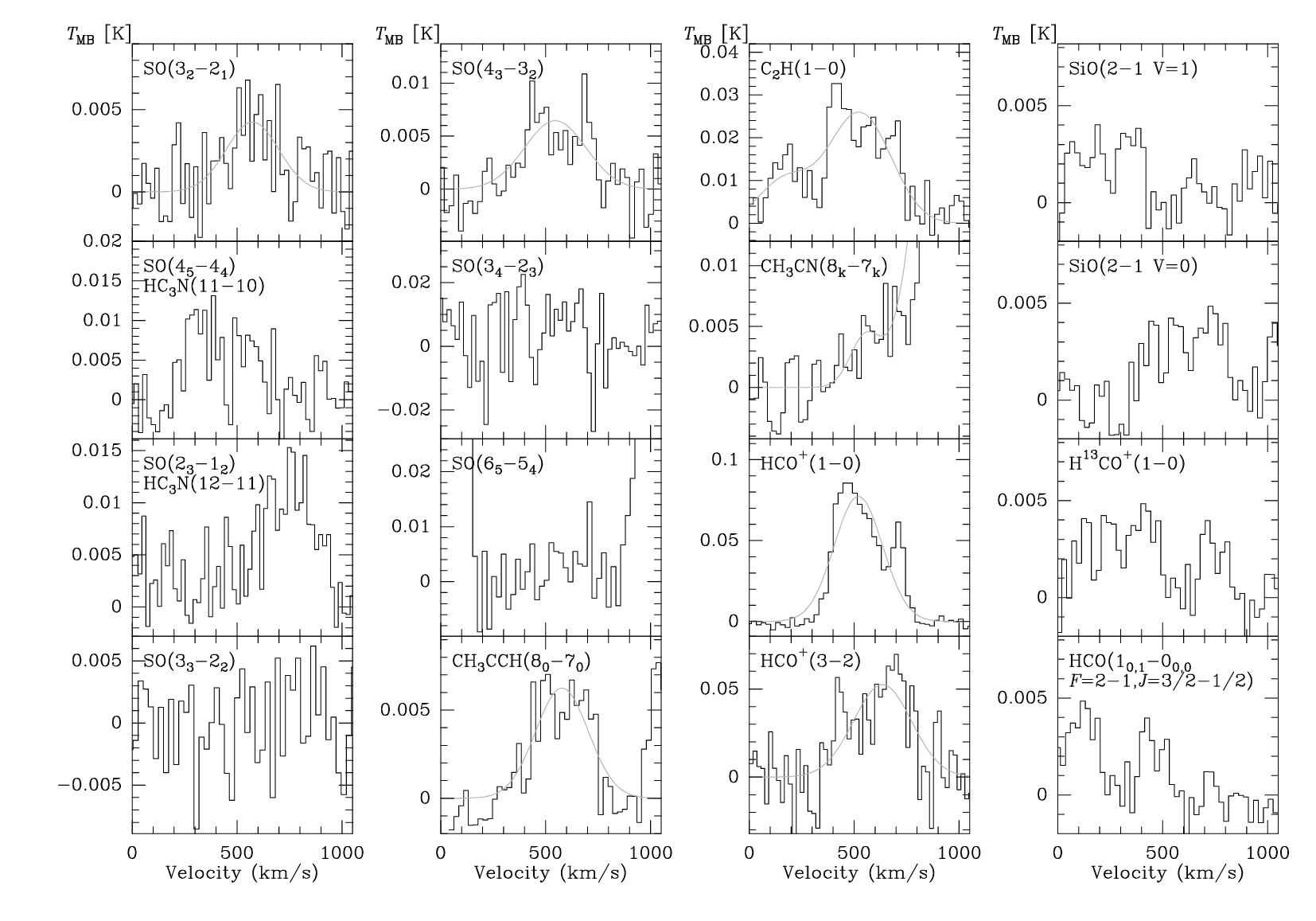}}
\caption[]{SO, CH$_3$CCH, C$_2$H, CH$_3$CN, HCO$^+$, SiO and HCO
spectra measured toward the center of NGC\,4945 (see also the
captions to Figs.\,1 and 2).}
\end{figure*}

\subsection{1.3~mm continuum measurements}

In July 1997, we mapped the 1.3\,mm continuum emission toward the
central region of NGC\,4945 with a single pixel bolometer at the
SEST. We first searched for extended emission by mapping a
$200''\times 200''$ area using the on the fly dual beam technique.
In addition, the inner $60''\times60''$ were mapped in raster mode
yielding a better signal-to-noise ratio.

Dual beam switching was performed using the SEST focal plane
chopping mirror with an offset of 67.5$''$ in azimuth and 3$''$ in
elevation.  Pointing was frequently checked on Cen\,A and on
NGC\,4945 itself. The mean pointing accuracy was $\pm 2.5''$.
Uranus was used as the primary calibrator with an assumed flux
density of 37\,Jy/beam. Fitting two dimensional (2D) Gaussians to
maps of Uranus yields a deconvolved full width to half power (FWHP)
beam size of $27.7''$. The sensitivity of the bolometer system was
determined to be 190\,mJy/beam\,s$^{-1}$.

The double beam maps were taken in a horizontal system with an
extent of $4'\times 3'$, a scanning velocity of 8$''$/s in
azimuth, and a step size in elevation of 8$''$. We took 36
individual maps, which were corrected for atmospheric extinction
using opacities obtained from skydips, and for the SEST
gain-elevation curve. Using the MOPSI package, the double beam
maps were converted into single-beam maps and transformed into
equatorial coordinates. The final coadded map has an rms of
30\,mJy/beam. The raster map was taken in asymmetric on-off mode
(off-on-on-on, 10\,s each phase) with an accumulated integration
time of 300\,s per point, thus reaching an rms of 11\,mJy/beam.
The 1.3\,mm  emission is slightly more extended than the beam
size. Fitting a 2D Gaussian to the on the fly map the FWHP extent
is $27.4''\times 30.8''$ with a position angle of 15$^{\rm o}$,
corresponding to a deconvolved size of $14''\times 20''$ or
$270\,{\rm pc} \times 390$\,pc. Both the on-the-fly map and the
raster map yield a maximum flux of 1.4\,Jy/beam including the line
emission. The flux density of the total emission mapped in our OTF
map is $2.8\pm 0.3$\,Jy of which $2.5\pm 0.3$\,Jy come from the
Gaussian shaped central emission. Again these values also include
spectral line emission. The separation of line and continuum flux
and a column density estimate are presented in Sect.~8.1.

\section{Results}

\subsection{The observed line profiles}

Figs.\,1--5 display the spectra from a total of 19 molecular
species that were measured toward the nuclear region of NGC\,4945,
nominally within 2$''$ of $\alpha = 13^{\rm h}05^{\rm m}27.2^{\rm
s}$, $\delta = -49^{\rm o}28' 05''$ (J2000.0). According to the
overall rotation curve (see Fig.\,5 in Mauersberger et al. 1996a),
blue-shifted line emission should arise southwest, red-shifted
emission should originate northeast of the dynamical center. For
lineshapes, see H94 (their Sect.\,5.2). Gaussian line parameters
to provide integrated intensities are given in Table~2, including
49 detected line features, 9 tentatively detected transitions and
22 undetected lines (for the notes to Table~2, see footnote
\footnote{ a) `+': detection; `$-$': non-detection; `?': tentative
detection `B': blended with
other lines, `N': no baseline. \\
b) Integrated from 300 to 800\,km\,s$^{-1}$  after subtracting a first order baseline
or a constant offset in $T_{\rm A}^{*}$. The errors were derived from Gaussian fits
and do not account for calibration and pointing uncertainties. \\
c) Obtained from single component Gaussian fits. ${v_{\rm LSR}}$ = ${v_{\rm HEL}}$ $-$
4.5\,km\,s$^{-1}$. \\
d) rms values for a $\sim $20\,km\,s$^{-1}$ channel width on a $T_{\rm mb}$ scale. \\
e) Channel spacings after smoothing as shown in Figs.\,1$-$5. \\
})

\begin{table*}
\begin{scriptsize}
\caption[]{Line parameters}
\begin{flushleft}
\begin{tabular}{l r c r @{$\pm$} l r @{$\pm$} r r @{$\pm$} r r c }

\hline

Transition & \multicolumn{1}{c}{Frequency} & Detection$^{\rm a)}$
     & \multicolumn{2}{c}{$\int$\,$T_{\rm mb}$\,d$v$$^{\rm b)}$}
     & \multicolumn{2}{c}{\vlsr $^{\rm c)}$} & \multicolumn{2}{c}{\delv $^{\rm c)}$}
     &\multicolumn{1}{c}{rms$^{\rm d)}$}
     &\multicolumn{1}{c}{$Dv$$^{\rm e)}$} \\
     & \multicolumn{1}{c}{(MHz)}
     &
     & \multicolumn{2}{c}{(K\,km\,s$^{-1}$)}
     & \multicolumn{2}{c}{(km\,s$^{-1}$)}
     & \multicolumn{2}{c}{(km\,s$^{-1}$)}
     & \multicolumn{1}{c}{(mK)} &\multicolumn{1}{c}{(km\,s$^{-1}$)} \\

\hline

HC$_3$N 9--8                                           &   81881.468     & + &    2.16     &   0.50       & 554.7    &  28.4      &  233.2  &   54.8  &  7.2  & 24.9  \\
C$_3$H$_2$ 2$_{1,2}$--1$_{0,1}$                          &   85338.905     & + &    2.71     &   0.26       & 515.8    &  21.8      &  459.5  &   48.2  &  1.9  & 23.9  \\
C$_3$H$_2$ 4$_{3,2}$--4$_{2,3}$                          &   85656.422     & + &   1.00      &   0.16       & 494.6    &  20.3      &  239.5  &   36.4  &  2.4  & 24.1  \\
HC$^{15}$N 1--0                                        &   86054.961     & + &   0.46     &   0.05       & 540.5    &  12.2      &  235.9  &   28.1  &  1.4   & 24.1  \\

SiO 2--1 $v$=1                                            &   86243.442       & $-$ &\multicolumn{2}{c}{...}&\multicolumn{2}{c}{...}&\multicolumn{2}{c}{...} & 1.9 & 24.1 \\
H$^{13}$CN 1--0                                        &   86340.184     & + &   0.96     &   0.08   & 551.1    &  12.8      &  301.2  &   21.6  &  1.0  & 23.6  \\
HCO 1$_{0,1}$--0$_{0,0}$, $J$=3/2--1/2                 &   86670.820     & B &\multicolumn{2}{l}{$\ \;$0.38 }    &\multicolumn{4}{l}{see Sect.\,5.3}  &  2.1  & 23.5  \\

H$^{13}$CO$^+$ 1--0                                    &   86754.294     & B &\multicolumn{2}{l}{$\ \;$0.59 }    &\multicolumn{4}{l}{see Sect.\,5.3}  &  2.1  & 23.5  \\
SiO 2--1 $v$=0                                              &   86846.998     & B &\multicolumn{2}{l}{$\ \;$0.74 }    &\multicolumn{4}{l}{see Sect.\,5.3}  &  2.1  & 23.5  \\

HN$^{13}$C 1--0                                        &   87090.942     & + &    1.06     &   0.13       & 615.0    &  21.8      &  358.0  &   48.8  &  1.5  & 23.5  \\
C$_2$H 1--0                                            &   87316.925     & + &    10.7     &   0.82       & 511.0    &  13.4      &  391.2  &   36.2  &  7.4  & 23.3  \\

HNCO 4$_{1,4}$--3$_{1,3}$                                &   87597.333     & $-$ &\multicolumn{2}{c}{...} &\multicolumn{2}{c}{...}&\multicolumn{2}{c}{...}& 1.9 & 23.3  \\
HNCO 4$_{0,4}$--3$_{0,3}$                                &   87925.238     & +&   2.83      &  0.55        & 575.0    & 27.5       &  289.6  &   64.0  &  4.5  & 23.2  \\
HCN 1--0                                               &   88631.847     & +&    22.4     &   0.40       & 549.7    &  2.8       &  304.8  &   5.6   &  10.5  & 23.0  \\
HCO$^+$ 1--0                                           &   89188.518     & +&   24.0      &  0.20        & 522.4    &  1.2       &  284.9  &   2.7   &  5.2  & 22.9  \\
HNC 1--0                                               &   90663.543     & +&    12.0     &   0.22       & 536.6    &  3.0       &  309.3  &   6.2   &  2.9  & 22.4  \\
HC$_3$N 10--9                                          &   90978.993     & + &    1.99     &  0.21        & 527.2    & 15.7       &  288.8  &   28.7  &  2.8  & 22.4  \\
N$_2$H$^+$ 1--0                                        &   93171.947     & +&   3.50      &  0.27        & 539.0    & 9.1        &  230.3  &   18.5  &  4.0  & 21.9  \\
CH$_3$OH 2$_1$--1$_1$ A+                               &   96396.010     & ? &   1.55      &  0.33        & 468.3    & 33.1       &  294.5  &   61.6  &  3.0  & 21.5  \\
CH$_3$OH 2--1                                          &   96741.420     & + &   6.00      &  0.38        & 576.8    & 7.8        &  245.6  &   17.6  &  3.6  & 21.5  \\

OCS 8--7                                               &   97301.208     & $-$     &\multicolumn{2}{c}{...} &\multicolumn{2}{c}{...}&\multicolumn{2}{c}{...} &1.9  & 21.0  \\
CS 2--1                                                &   97980.968     & +&   16.9      &  0.91        & 564.8    & 9.7        &  378.6  &   25.5  &  6.6  & 20.8  \\
SO 3$_2$--2$_1$                                        &   99299.879     & + &    1.24     &  0.28        & 568.9    &  25.9      &  259.9  &   75.6  &  2.6  & 20.5  \\
SO 4$_5$--4$_4$                                        &  100029.569     & B &\multicolumn{2}{c}{...} &\multicolumn{2}{c}{...}&\multicolumn{2}{c}{...}& 1.9 & 20.4  \\
HC$_3$N 11--10                                         &  100076.389     & + &     2.92    &  0.35        & 542.9    &   23.2     &  341.2  &   42.7  &  4.6  & 20.4  \\
$^{13}$CN 1--0, $J$=1/2--1/2                              &  108657.646     & $-$ &\multicolumn{2}{c}{...}  &\multicolumn{2}{c}{...}&\multicolumn{2}{c}{...}& 1.9 & 18.8  \\
$^{13}$CN 1--0, $J$=3/2--1/2                              &  108780.201     & $-$ &\multicolumn{2}{c}{...} &\multicolumn{2}{c}{...}&\multicolumn{2}{c}{...}& 1.9 & 18.8  \\
HC$_3$N 12--11                                         &  109173.634     & +&     4.18    &  0.38        & 521.7    &   15.1     &  341.1  &   39.0  &  4.5  & 18.7  \\

SO 2$_3$--1$_2$                                        &  109252.184     & B&\multicolumn{2}{c}{...} &\multicolumn{2}{c}{...}&\multicolumn{2}{c}{...}& 1.9 & 18.7  \\
OCS 9--8                                               &  109463.063     & ?&   2.93     &  0.67        & 669.9    &  34.7      &  483.9  &   122   &    4.3 & 18.7  \\
C$^{18}$O 1--0                                         &  109782.160     & +&   9.29     &  0.42        & 539.2    &   6.4      &  274.8  &   12.5  &  12.9  & 18.6  \\
HNCO 5$_{0,5}$--4$_{0,4}$                                &  109905.753     & ?&     1.13    &    0.43      &  478.7   &   44.5     &   200.3 &    62.3 &   5.5 &  18.6  \\
$^{13}$CO 1--0                                         &  110201.353     & +&     34.4    &     1.2      &  538.4   &   4.5      &   246.4 &    8.7  &   19.5 &  18.6  \\
C$^{17}$O 1--0                                         &  112358.780     & +&     1.21    &    0.23      &  596.8   &   30.4     &   297.3 &    54.5 &   2.1 & 36.4  \\
CN 1--0, $J$=1/2--1/2                                  &  113191.317     & +&     15.5    &    0.5       &  405.9   &   5.8      &   351.5 &    11.8 &   6.7 & 18.0  \\
CN 1--0, $J$=3/2--1/2                                  &  113490.982     & +&     24.5    &    0.5       &  499.3   &   3.4      &   304.7 &    8.1  &   7.6 & 18.0  \\
CO 1--0                                                &  115271.204     & +&     523.1   &    1.5       &  545.5   &   0.4      &   268.8 &    0.9  &  50.8 & 17.7  \\

SO 3$_3$-2$_2$                                        &  129138.898     & $-$ &\multicolumn{2}{c}{...} &\multicolumn{2}{c}{...}&\multicolumn{2}{c}{...}& 1.9 & 24.2  \\
HNCO 6$_{0,6}$--5$_{0,5}$                                &  131885.740     & +&     2.93    &    0.18      &  570.0   & 8.8        &   284.1 &    17.6 &   2.9 & 23.7  \\
HC$_3$N 15--14                                         &  136464.400     & + &     2.13    &    0.29      &  582.2   & 18.9       &   249.0 &    32.2 &   5.4 & 22.9  \\
CH$_3$CCH 8$_0$--7$_0$                                 &  136728.010     & + &     1.77    &    0.23      &  576.7   & 18.9       &   280.2 &    35.8 &   3.9 & 22.9  \\
SO 4$_3$--3$_2$                                        &  138178.648     & +&     2.41    &    0.15      &  548.0   &  11.2      &   335.1 &    21.5 &   2.4 & 22.6  \\
$^{13}$CS 3--2                                         &  138739.309     & ?&    0.28   &    0.10  &  547.8   & 51.9       &   289.2 &   93.3  &   8.9 & 22.5  \\
H$_2$CO 2$_{1,2}$--1$_{1,1}$                             &  140839.518     & + &     3.73    &    0.28      &  535.0   &  9.5       &   241.4 &    18.6 &   5.1 & 22.2  \\
C$^{34}$S 3--2                                         &  144617.147     & +  &   1.34    &    0.10      &  574.2   &  11.4  &   298.3 &   23.8  &  1.1  & 21.7  \\
DCN 2--1                                               &  144828.000     & N&\multicolumn{2}{c}{...}&\multicolumn{2}{c}{...}&\multicolumn{2}{c}{...}& 1.9 & 21.6  \\
CH$_3$OH 3--2                                          &  145124.410     & +&    9.36     &   0.06   &  603.3   & 0.9        &  268.7  &   2.0   &  2.3  & 21.7  \\
HC$_3$N 16--15                                         &  145560.946     & + &     $<$5.02     &   0.19       &  518.8   &  6.2       &  327.6  &   13.5  &  3.0  & 21.4  \\  

H$_2$CO 2$_{0,2}$--1$_{0,1}$                             &  145602.953     & + &   $<$5.02     &   0.19       &  605.7   &  6.2       &  327.6  &   13.5  &  3.0  & 21.4  \\
CS 3--2                                                &  146969.049     & + &     13.9    &    0.2       &  549.4   &   2.2      &   311.5 &    4.1  &   5.8 & 21.4  \\
CH$_3$CN 8$_k$--7$_k$                                  &  147174.592     & ? & 0.77        &    0.12      &  555.2   &   23.4     &   173.5 &    40.0 &   2.0   & 21.2  \\
CH$_3$OH 14$_0$--14$_{-1}$ E                           &  150141.680     & ? &    1.65     &   0.28       &  460.6   &  27.7      &  311.5  &   54.2  &  4.7   & 20.7  \\
H$_2$CO 2$_{1,1}$--1$_{1,0}$                           &  150498.339     & +&    5.78     &   0.19       &  535.1   &  5.4       &  320.9  &   11.4  &  3.1  & 20.7  \\
C$_3$H$_2$ 4$_{0,4}$--3$_{1,3}$ \& 4$_{1,4}$--3$_{0,3}$     &  150835.000     & +&    1.79     &   0.22       &  576.4   &  16.7      &  259.5  &   30.2  &  4.1  & 20.7  \\
HNCO 7$_{0,7}$--6$_{0,6}$                                &  153865.092     & +&    6.23     &   0.39       &  589.8   & 10.2       &  352.4  &   29.3  &  2.9  & 20.3  \\
HC$_3$N 17--16                                         &  154657.283     & ? &    2.26     &    0.55      &  525.8   & 36.1       &   282.5 &    76.3 &   4.9 & 20.2  \\
SO 3$_4$--2$_3$                                        &  158971.814     & $-$ &\multicolumn{2}{c}{...} &\multicolumn{2}{c}{...}&\multicolumn{2}{c}{...}& 1.9 & 19.6  \\
H$_2$CO 3$_{0,3}$--2$_{0,2}$                             &  218222.191     & $-$ &\multicolumn{2}{c}{...} &\multicolumn{2}{c}{...}&\multicolumn{2}{c}{...}& 1.9 & 19.0  \\
HC$_3$N 24--23                                         &  218324.744     & $-$ &\multicolumn{2}{c}{...} &\multicolumn{2}{c}{...}&\multicolumn{2}{c}{...}& 1.9 & 22.9  \\
C$^{18}$O 2--1                                         &  219560.319     & +&      24.5   &   0.5        &  585.6   &  3.1       &  287.3  &    6.1  &  10.6  & 22.7  \\
HNCO 10$_{0,10}$--9$_{0,9}$                            &  219798.282     & $-$ &\multicolumn{2}{c}{...} &\multicolumn{2}{c}{...}&\multicolumn{2}{c}{...}& 1.9 & 22.7  \\
SO 6$_5$--5$_4$                                        &  219949.391     & $-$ &\multicolumn{2}{c}{...} &\multicolumn{2}{c}{...}&\multicolumn{2}{c}{...}& 1.9 & 18.9  \\
$^{13}$CO 2--1                                         &  220398.686     & +&      81.2   &   0.7        &  568.1   &  1.3       &  287.3  &    2.5  &  15.1  & 22.6  \\
C$^{17}$O 2--1                                         &  224714.368     & + &      3.31   &   0.31       &  611.4   &   12.3     &  252.7  &    25.6 &  4.1  & 22.2  \\
H$_2$CO 3$_{1,2}$--2$_{1,1}$                             &  225697.772     & ?&      1.73   &   0.46       &  567.3   &  36.0      &  247.9  &    60.2 &  5.6  & 22.1  \\
CN 2--1                                                   &  226874.764     & + &   29.4     &    0.39      &  638.2  &   18.9      &  343.3  &    33.3  & 18.1  & 22.0 \\
CH$_3$OH 21$_1$--20$_0$  E                             &  227094.600     & $-$ &\multicolumn{2}{c}{...} &\multicolumn{2}{c}{...}&\multicolumn{2}{c}{...}& 1.9 & 22.1  \\
HC$_3$N 25--24                                         &  227418.957     & $-$ &\multicolumn{2}{c}{...} &\multicolumn{2}{c}{...}&\multicolumn{2}{c}{...}& 1.9 & 22.1  \\
CO 2--1                                                &  230538.000     & +&     920.9   &    0.6       &  555.8   &  0.1   &   275.1 &   0.2   &   25.6 & 21.6  \\
CH$_3$OH 5--4                                          &  241767.224     & + &    $<$6.80    &    0.88      &  582.5   &   21.8     &   325.0 &    43.7 &   6.5 & 20.6  \\ 
HNCO 11$_{0,11}$--10$_{0,10}$                          &  241774.037     & B &    $<$6.80    &    0.88      &  591.0   &   21.8     &   325.0 &    43.7 &   6.5 & 20.6  \\
CH$_3$OH 5$_1$--4$_1$,A-                               &  243915.826     & $-$ &\multicolumn{2}{c}{...} &\multicolumn{2}{c}{...}&\multicolumn{2}{c}{...}& 1.9 & 17.0  \\
CS 5--4                                                &  244935.606     & + &     9.98    &    0.67      &  582.8   &   9.9      &   286.1 &    20.3 &   6.0 & 20.5  \\
CH$_3$OH 21$_{-3}$--21$_{-2}$ A                        &  245223.000     & $-$ &\multicolumn{2}{c}{...} &\multicolumn{2}{c}{...}&\multicolumn{2}{c}{...}& 1.9 & 20.4  \\
H$^{13}$CN 3--2                                        &  259011.790     & + &   0.98     &   0.17       &  543.3   &  21.7      &   250.7 &     46.9 &   8.1 & 19.4  \\
HCN 3--2                                               &  265886.432     & +&    29.6     &   0.7        &  592.6   &   4.1      &   352.6 &    8.1  &   27.9 & 18.8  \\
HCO$^+$ 3--2                                           &  267557.625     & +&     18.3    &     1.9      &  635.2   &   17.0     &   330.7 &    37.8 &   20.3 & 18.6  \\
$^{13}$CO 3--2                                         &  330587.957     & + &    87.9     &    9.6       &  499.9   &   17.8     &   304.0 &    44.4 &  135      & 21.5  \\
CO 3--2                                                &  345795.991     & + &     870.6   &     7.7      &  495.9   &  1.0       &   226.8 &    2.5  &   211     & 20.6  \\
HCN 4--3                                               &  354505.472     & ? &    7.18     &    1.74      &  627.9   &   50.5     &   400.9 &    87.2 &  14.1  & 20.1  \\
\hline

\end{tabular}
\end{flushleft}
\end{scriptsize}
\end{table*}

\subsection{Line intensities}

In order to compare our observational data with previously published results, we list
in Table~3 integrated line intensities ($\int$$T_{\rm mb}$\,d$v$) and peak line
temperatures ($T_{\rm mb}$). In columns 2 and 5, we display integrated intensities and
peak line temperatures derived from our data. In columns 3, 4, 6 and 7, data from
C01, H94 and Mauersberger et al. (1996a) are given. There is good agreement for the
strong CO lines, i.e. CO and $^{13}$CO. Differences are 20\% or less for the $J$=1--0
and 2--1 transitions. Not surprisingly, for the weaker lines, differences are larger.
This is likely a consequence of lower signal-to-noise ratios and a lower pointing
quality, caused by pointing shifts during one hour long integrations and the fact that
checks of the shape of the profile, showing whether the pointing was stable or not,
were not possible for individual scans. Because the CO and $^{13}$CO spectra encompass
the entire frequency range of our observations, we conclude that serious systematic
calibration problems are absent in H94, Mauersberger et al. (1996a), C01 and our new
data. Significant differences in line intensity and lineshape that are seen in the weaker
lines are interpreted as the simultaneous consequence of noise and pointing errors and
will be analysed in the following sections.

\begin{table}
\begin{threeparttable}
\begin{scriptsize}
\begin{minipage}{180mm}
\caption[]{{A comparison of line intensities with C01, H94 and Mauersberger et al. (1996a)}}
\begin{flushleft}
\begin{tabular}{lccclll}
\hline\hline
line   & \multicolumn{3}{l}{$\int$$T_{\rm mb}$\,d$v$}
       &\multicolumn{3}{l}{$T_{\rm mb}$} \\ 
       & \multicolumn{3}{l}{(K km s$^{-1}$)}  
       & \multicolumn{3}{l}{(K)} \\   
\hline
(1)\tnote{a} &
(2)\tnote{b} &
(3)\tnote{c} &
(4)\tnote{d} &
(5)\tnote{b} &
(6)\tnote{c} &
(7)\tnote{d} \\
\hline\hline
CO 1--0                       & 523     &  510       &         466\tnote{e} &     1.8    &   1.7\tnote{e}   &1.6\tnote{e}                  \\
CO 2--1                       & 921     &  740       &        1050 &     3.2    &   2.4   &3.2                   \\
CO 3--2                       & 871     &  760       &             &     3.5    &   1.6   &2.5\tnote{f}        \\
$^{13}$CO 1--0                & 34.4    &  30        &         32.3&     0.12   &   0.114 &0.107                 \\
$^{13}$CO 2--1                & 81.2    &  86        &        123.2&     0.24   &   0.37  &0.4                   \\
$^{13}$CO 3--2                & 88.0    &            &             &     0.30   &         &0.25\tnote{f}        \\
C$^{18}$O 1--0                & 9.29    &  8.4       &        10.9 &     0.03   &   0.026 &0.037                 \\
C$^{18}$O 2--1                & 24.5    &  29        &         32.4&     0.07   &   0.12  &0.12                  \\
C$^{17}$O 1--0                & 1.21    &  1.8       &         1.33&     0.004  &   0.005 &0.0057               \\
C$^{17}$O 2--1                & 3.31    &  5         &             &     0.012  &   0.019 &\\
CS 2--1                       & 17.0    &  9.3       &             &     0.04   &   0.031 &\\
CS 3--2                       & 13.9    &  11        &             &     0.038  &   0.038 &\\
HCN 1--0                      & 22.4    &  24        &         23.5&     0.065  &   0.08  & 0.08   \\
HCN 3--2                      & 29.6    &  45        &             &     0.09   &   0.2   &\\
HCO$^+$ 1--0                  & 24.0    &  21        &             &     0.08   &   0.08  &\\
HCO$^+$ 3--2                  & 18.3    &  37        &             &     0.06   &   0.225 &\\
H$_2$CO 2$_{1,2}$--1$_{1,1}$  & 3.73    &  6.5       &             &     0.013  &   0.02  &\\
H$_2$CO 2$_{0,2}$--1$_{0,1}$  & 5.02    &  5.4       &             &     0.014  &   0.017 &\\
H$_2$CO 2$_{1,1}$--1$_{1,0}$  & 5.78    &  5.9       &             &     0.019  &   0.013 &\\
\hline\hline
\end{tabular}
\begin{tablenotes}
\item[a](1) lines selected from C01 except $^{13}$CO 3--2, which is from Mauersberger et al. 1996a;
\item[b](2) \& (5) our data;
\item[c](3) \& (6) C01;
\item[d](4) \& (7) H94;\\
\item[e] applied beam efficiencies for columns (4), (6) and (7):
0.75 for HCN\,1--0 and HCO$^+$\,1--0; 0.70 for CO\,1--0,
$^{13}$CO\,1--0, C$^{18}$O\,1--0, C$^{17}$O\,1--0 and CS\,2--1;
0.60 for CS\,3--2 and the three H$_2$CO lines; 0.50 for CO\,2--1,
$^{13}$CO\,2--1, C$^{18}$O\,2--1, and C$^{17}$O\,2--1; 0.40 for
CO\,3--2, $^{13}$CO\,3--2, HCN\,3--2 and HCO$^+$\,3--2;\\
\item[f] Mauersberger et al. 1996a
\end{tablenotes}
\end{flushleft}
\end{minipage}
\end{scriptsize}
\end{threeparttable}
\end{table}

\subsection{Extent of the survey}

The number of measured lines toward NGC\,4945 is significantly larger than that
in previous studies (Henkel et al. 1990; H94; C01). While the most complete study
carried out so far, that of C01, contains a comparable number of lines from CO and
its isotopomers as well as from H$_2$CO, the number of CS and SO lines studied in
our survey is larger. About 51\% of the frequency range between 81.4\,GHz and 115.8\,GHz,
42\% of the frequency range between 128.6\,GHz and 159.5\,GHz, and 27\% of the frequency
range between 217.7\,GHz and 268.1\,GHz are covered. Of particular importance is
HC$_3$N which was not clearly detected in any previous study (see H94 for a tentative
detection) but which was detected in our survey in at least half a dozen of lines.
This molecule is useful to determine H$_2$ densities (e.g. Mauersberger et al. 1990)
and can be used to check previously determined physical parameters of molecular clouds
(e.g. C01). The (tentative) detection of a number of additional molecular species,
like CH$_3$CCH and CH$_3$CN provides a broader view of chemical abundances while
the detection of  C$^{34}$S and the non-detection of $^{13}$CN give new insights into
isotopic ratios in the nuclear environment of NGC\,4945 (see Henkel \& Mauersberger
1993 and C01 for earlier summaries on extragalactic CNO isotope ratios, including
NGC\,4945).

\section{Radiative transfer calculations}

With the peak and integrated line intensities determined from our measured profiles,
we can derive a number of relevant line intensity ratios that can be used to estimate
density, column density and kinetic temperature. The excitation problem involves
statistical equilibrium of a multilevel system that was solved in most cases by the
Large Velocity Gradient (LVG) method. This allows us to determine a local source
function in which optical depths are related to escape probabilities of the photons.
The model gas cloud has spherical geometry and uniform kinetic temperature and density.
While there may be molecules which are significantly affected by the intense
radiation field in the nuclear region of NGC\,4945 (for effects on HNCO, see
Churchwell et al. 1986 and Sect.\,4.9; for the negligible effects on CO, see
Mao et al. 2000 and their Appendix C), such effects are difficult to quantify.
If not said otherwise, a background temperature of 2.73\,K is therefore used
throughout the discussion.

With lineshapes being approximately similar (for a discussion of
the moderate differences obtained, see e.g. Bergman et al. 1992
and H94), measured peak intensities (i.e. not the peak intensities
obtained from Gaussian fits) appear to be more reliable input
parameters for the model than integrated intensities (see
Sect.\,4.1). Taking CO source sizes according to Table~4 of C01,
i.e. 29$''$ for the $J$=1--0, 20$''$ for the 2--1 and 15$''$ for
the 3--2 transition, we corrected for beam dilution effects by
calculating $T^{ \prime}_{\rm mb}$=$T_{\rm mb}$/$\eta_{\rm bf}$,
with $\eta_{\rm bf}$ = $\theta^2_{\rm s}$/($\theta^2_{\rm
s}$+$\theta^2_{\rm b}$), $\theta_{\rm s}$ and $\theta_{\rm b}$
denoting the Gaussian FWHP source and beam size, respectively.
Again following C01, for the other molecular species a source size
of 20$''$ was adopted. This appears to be a reasonable guess in
view of the CO data and the extent of the 1.3\,mm continuum
emission (see Sect.\,2.2). Table~4 gives beam dilution corrected
line temperatures, which are used for the LVG calculations.

Results of the model calculations are shown below, displaying line
intensities and line intensity ratios as a function of H$_2$
density and molecular column density. Common to all these plots
are source size averaged line temperatures. Cloud averaged line
temperatures may be significantly larger, depending on the degree
of clumping. To give an example: if the HCN $J$=1--0 emission
arises from the innermost 20$''$ of the galaxy and if we assume a
rotation temperature of $T_{\rm rot}$=10\,K and $\tau$\,(HCN
1--0)$\geq$\,1 (H94), the `clumping factor' within the 20$''$
sized region would become $f_{\rm c}\sim$\,0.1. As long as the
lines are optically thin, this does not imply a change in
excitation or in the line temperature ratios. If, however, the
transition from source to cloud averaged quantities involves
optical depths near unity or higher, significant changes are
expected.

In the following, LVG simulations are discussed for some of the
most important molecular species, carefully discriminating between
optically thick and optically thin line emission. Most
calculations were made for kinetic temperatures of 50 and 100\,K.
Such high values are justified as e.g. suggested by the multilevel
CO studies of Mao et al. (2000) and Bradford et al. (2003) on
M\,82 and NGC\,253, respectively, and by the NH$_3$ study of
Mauersberger et al. (2003) towards NGC\,253, Mafffei 2, IC\,342,
and M\,82. As we shall see, even for $T_{\rm kin}$ = 100\,K H$_2$
densities are larger than previously anticipated (c.f. C01 and
Sect.\,7) and would have to be further enhanced if lower $T_{\rm
kin}$ values would be adopted.

\subsection{CO}

Plotting CO/$^{13}$CO $J$=1--0 and 2--1 line intensity ratios as a
function of velocity (Fig.\,6), we obtain maxima and minima that
agree with those shown by Bergman et al. (1992; their Figs.\,2 and
3) for the $J$=1--0 lines alone. This is consistent with the
hypothesized ring-like morphology of the emitting gas (see also
Whiteoak et al. 1990; Dahlem et al. 1993; Mauersberger et al.
1996a).

\begin{figure}[ht]
\begin{minipage}[t]{8.5cm}
\resizebox{\hsize}{!}{\includegraphics[angle=0]{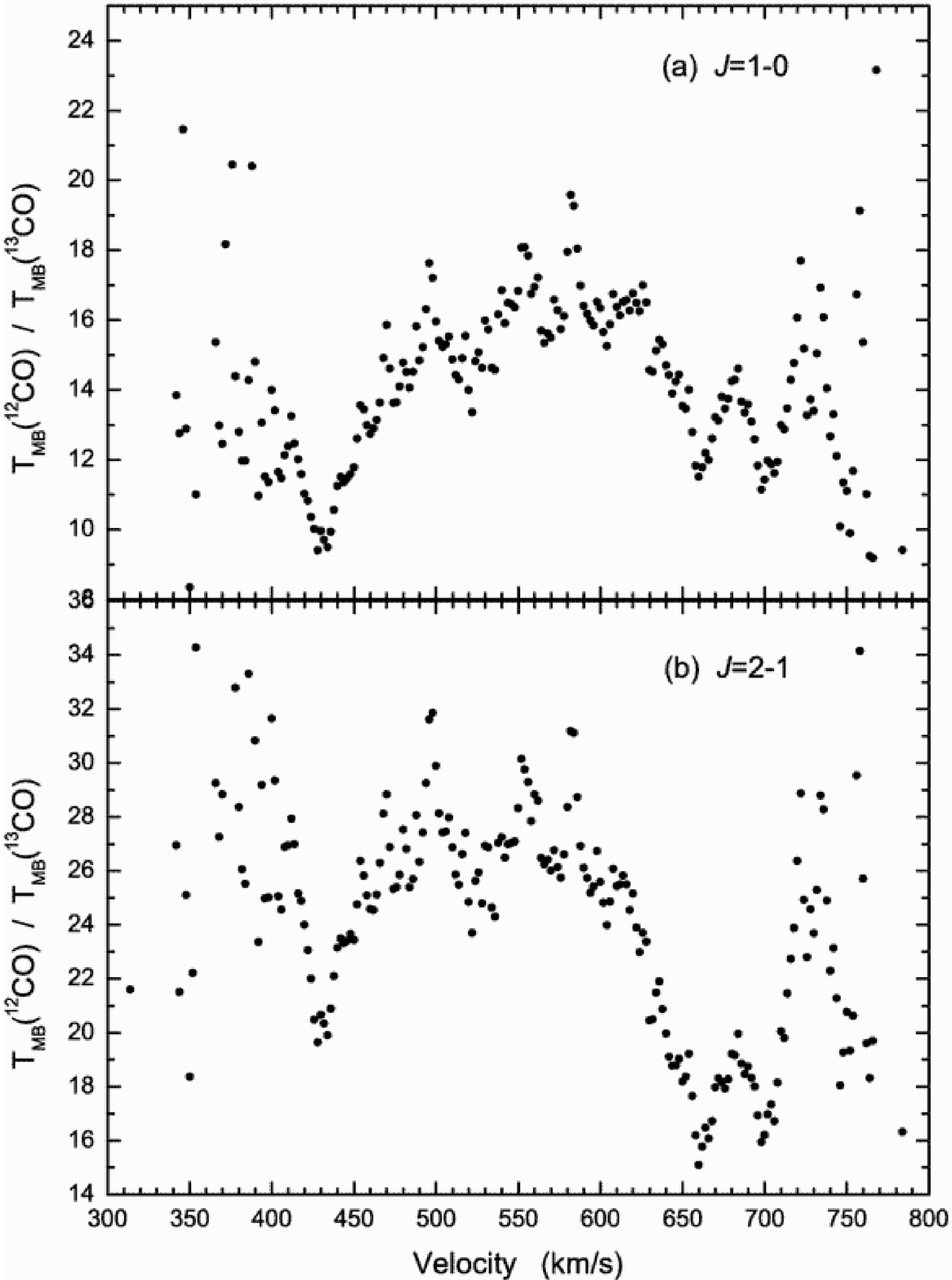}}
\end{minipage}
\caption[]{CO/$^{13}$CO intensity ratios as a function of radial velocity for the
$J$=1--0 (upper panel) and $J$=2--1 (lower panel) lines.}
\end{figure}

Among the 10 detected lines of carbon monoxide (CO; Fig.\,1, first
three panels), the C$^{17}$O $J$=1--0 and the two $J$=3--2
profiles are the most uncertain. The signal-to-noise ratio of the
C$^{17}$O $J$=1--0 line is small, but the line intensity is
similar to that of the higher signal-to-noise detection reported
by C01. Our C$^{17}$O $J$=2--1 profile is of better quality. With
respect to the CO and $^{13}$CO $J$=3--2 profiles, that were not
measured simultaneously, we note that the lineshape is narrow,
mainly confined to the lower radial velocity range of the CO 1--0
and 2--1 profiles. Since we know from Mauersberger et al. (1996a)
that CO $J$=3--2 lines have no peculiar shape, the effect must be
due to small pointing errors. These can substantially affect our
submillimeter data that were obtained with a relatively small
beam. The CO and $^{13}$CO $J$=3--2 profiles, while not showing
emission over the entire velocity range, likely indicate the
correct peak line temperatures, since the $J$=1--0 and 2--1 line
peaks are broad ($>$\,100\,km\,s$^{-1}$) and since the peak
temperatures are not lower than those reported by Mauersberger et
al. (1996a) and C01. This is the major motivation to use peak and
not integrated line intensities for the model calculations.

Applying an LVG code to the CO data (see Mao et al. 2000) and
reproducing C$^{17}$O and C$^{18}$O $J$=1--0 and 2--1 lines, we
obtain an excitation parameter of $T_{\rm kin}\cdot n^{3/2}_{\rm
H_2}$\,$\sim$\,2$\times$10$^7$\,K\,cm$^{-9/2}$, i.e. a kinetic
temperature of $T_{\rm kin}$\,$\sim$\,100\,K yields a density of
$n_{\rm H_2}$\,$\sim$\,3.5$\times$10$^3$\,cm$^{-3}$, in good
agreement with C01 (their Table~5). Also the resulting column
densities
$$
N=[X^{\prime}/(dv/dr)]\times3.08\times10^{18}\times\Delta v_{1/2}
$$
($N$: column density in cm$^{-2}$; $X^{ \prime}$: LVG density
parameter of the specifically studied molecule in cm$^{-3}$;
$dv/dr$: velocity gradient in km\,s$^{-1}$\,pc$^{-1}$; $\Delta
v_{1/2}$: linewidth in km\,s$^{-1}$, the adopted value is
300\,km\,s$^{-1}$), show with 3.7$\times$10$^{15}$\,cm$^{-2}$ and
2.4$\times$10$^{16}$\,cm$^{-2}$ for C$^{17}$O and C$^{18}$O,
respectively, good agreement. If we consider errors in individual
line temperatures by $\pm$10\%, we obtain
$N$(C$^{18}$O)/$N$(C$^{17}$O)=6.4$\pm$0.3.

Fits to the $^{13}$CO and CO peak line intensities, involving three transitions
in each case, are more difficult to interpret. Using the temperature and density
deduced above implies 2--1/1--0 and 2--1/3--2 line intensity ratios that are larger
than those observed. A $^{13}$CO column density of 9.2$\times$10$^{16}$ cm$^{-2}$,
that approximately reproduces observed line intensities, leads to 2--1/1--0 and
2--1/3--2 line intensity ratios of 1.8 and 1.6, respectively, that have to be compared
with observed values of 1.3 and 0.9. Assuming that the 2--1 line, that has the highest
optical depth, arises from a particularly large volume, makes the discrepancy even more
pronounced (note that this is independent of the overall extent of the emission that is
given in Table~4 of C01; it refers instead to the source filling or clumping factor
within that volume). The same holds for CO, where a model fit with $N$(CO) =
1.8$\times$10$^{18}$\,cm$^{-2}$ yields 2--1/1--0 and 2--1/3--2 ratios of 1.6 and 1.3,
respectively, instead of the observed ratios of 1.2 and 1.0. Furthermore, an abundance
ratio of $\sim$\,20 between CO and $^{13}$CO may be too small in view of a $^{12}$C/$^{13}$C
ratio of $\sim$\,50, that was suggested as a likely value by H94 and C01.

What can resolve the obtained discrepancies? As already mentioned,
$J$=3--2 peak line temperatures may be particularly uncertain.
Here calibration errors can be more significant than for the other
lines. Our CO $J$=3--2 peak intensity is larger by 30\% than that
reported by C01. Taking their spectrum would lead to a reasonable
model fit. The shape of our $^{13}$CO $J$=3--2 spectrum (Fig.\,1)
is unusual. It remains open whether the narrow feature at
$\sim$\,430 km\,s$^{-1}$ ($T_{\rm mb}$\,$\sim$\,0.4\,K) or the
wider component at higher velocities ($T_{\rm
mb}$\,$\sim$\,0.25\,K) represents the proper peak intensity value.
A good match with the model would imply $T_{\rm
mb}$\,$\sim$\,0.12\,K. Thus there could be a serious discrepancy
between the model and the observational data that has to be
addressed by future studies.

Measured 2--1/1--0 line intensity ratios are more accurate. The CO $J$=2--1 line of
C01 is weaker than the corresponding line measured by us and would enhance the conflict
between the model and observations. Having observed the line on many occasions, however,
we believe that our calibration (and that of H94) is more appropriate. Otherwise,
differences with previously reported $J$=1--0 and 2--1 intensities are negligible.

One way to reconcile observed line intensities with those of the
model is to assume clumping. As long as all line emission is
optically thin, intensities will be proportional to $f_{\rm
c}\times\tau$, i.e. to the product of the clumping factor (the
line emitting area ${\it within}$ the cloud size given in
Sect.\,4) and optical depth. For a given line intensity a decrease
in the clumping factor $f_{\rm c}$ would then imply a
corresponding increase in optical depth. If $\tau$ approaches
unity, however, $f_{\rm c}\times\tau$ does not determine line
intensities any longer and each of the factors has to be analyzed
separately. The C$^{17}$O, C$^{18}$O, $^{13}$CO and CO column
densities were derived assuming $f_{\rm c}$=1, reducing all
optical depths to their possible minimum. For CO, however, this
approach is too simple. If $f_{\rm c}$ is identical for all CO
isotopomers and $f_{\rm c}<1$, optical depths and column densities
will be larger than the so far calculated values ignoring clumping
within the CO emission region. Higher opacities imply more photon
trapping and thermalization, thus reducing differences between
$J$=1--0, 2--1 and 3--2 line intensities and leading to a better
agreement with observed line temperature ratios. As already
mentioned, homogeneous cloud coverage yields
$N$(CO)\,${\sim\,2\times 10^{18}}$\,cm$^{-2}$ (Fig.\,7a,b),
$N$(CO)/$N$($^{13}$CO)\,$\sim$\,20, and 2--1/1--0 and 2--1/3--2
line intensity ratios of 1.6 and 1.3. A cloud averaged column
density of $N$(CO) $\sim$ 6$\times$10$^{18}$ cm$^{-2}$ implies
$N$(CO)/$N$($^{13}$CO)\,$\sim$\,50 ($f_{\rm c} \sim 0.4$; see
Fig.\,7c) and line intensity ratios of 1.4 and 1.2, that agree
better with the observed ratios of 1.2 and 1.0, respectively.

While moderate (0.2$<f_{\rm c}<$1) clumping is thus likely, the procedure is not
accurate enough to provide reliable values for $f_{\rm c}$, $N$(CO) and $N$(CO)/$N$($^{13}$CO).
In the following subsections, we will therefore try to constrain the $^{12}$C/$^{13}$C
isotope ratio in other ways. Assuming a constant carbon isotope ratio throughout the
nuclear region, this will yield the proper $N$(CO)/$N$($^{13}$CO) abundance and
$^{12}$C/$^{13}$C isotope ratios. A comparison with the beam averaged CO column density
will then also determine the clumping factor.

\begin{figure}[ht]
\begin{minipage}[t]{24.0cm}
\resizebox{\hsize}{!}{\includegraphics[angle=-90]{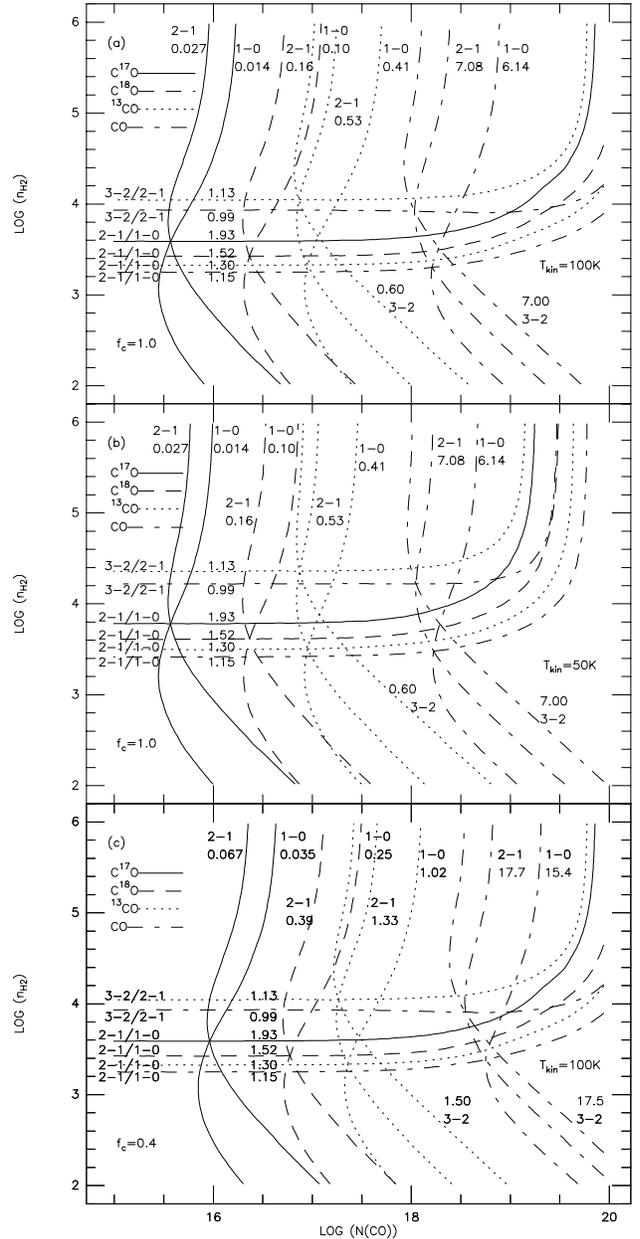}}
\end{minipage}
\caption[]{LVG calculations with the H$_2$ density $n_{\rm H_2}$
(in cm$^{-3}$) as a function of column density $N$ (in cm$^{-2}$)
for CO (dash-dotted lines) and its isotopomers C$^{17}$O (solid
lines), C$^{18}$O (dashed lines) and $^{13}$CO (dotted lines) at
$T_{\rm kin}$\,=\,100\,K and $f_{\rm c}$=1.0 (a), $T_{\rm
kin}$\,=\,50\,K and $f_{\rm c}$=1.0 (b) and $T_{\rm
kin}$\,=\,100\,K and $f_{\rm c}$=0.4 (c). $f_{\rm c}$ is the
clumping factor (see Sect.\,4.1). Transitions and brightness temperatures
are given for each of the `vertical' lines. The `horizontal' lines
connect points with a specific line intensity ratio in the ($N$,
$n_{\rm H_2}$) plane.}
\end{figure}

\subsection{CS}

Unlike CO with its high column density, optical depth and small critical density
(where rates of collisional and radiative deexcitation become similar), carbon
monosulfide (CS) is characterized by smaller column densities and larger critical
densities. While CO may arise predominantly from warm layers at the surface of clouds
that are irradiated by a strong UV radiation field (e.g. Mao et al. 2000), CS emission
likely originates from denser clumps of gas deeper inside the clouds where kinetic
temperatures might be lower. While the study of C01 was confined to the CS $J$=2--1
and 3--2 lines, our LVG model fit (see Mauersberger \& Henkel 1989) also includes
the $J$=5--4 line that is crucial for an excitation analysis as well as the $J$=3--2
lines of two rare isotopomers, those of C$^{34}$S and $^{13}$CS.

At a kinetic temperature of $T_{\rm kin}$\,$\sim$\,100\,K, the $J$=2--1, 3--2, and
5--4 line temperatures lead to $N$(CS)\,$\sim$\,4$\times$10$^{14}$\,cm$^{-2}$ and
$n_{\rm H_2}$\,$\sim$\,5$\times$10$^4$\,cm$^{-3}$ (see Fig.\,8). The CS column
density is smaller, the H$_2$ density is larger than reported by C01, by factors
of five. Assuming a lower kinetic temperature does not significantly alter $N$(CS),
but increases the density further. At $T_{\rm kin}$=25\,K, $n_{\rm H_2}$ would be
in excess of 10$^5$\,cm$^{-3}$.  The reason for the discrepancy between C01 and
our data lies in the fits to the $J$=2--1 and $J$=5--4 lines. With the parameters
suggested by C01, the $J$=2--1 line would be stronger and the $J$=5--4 line would
be weaker than observed.

The model calculations of C01 and our results both indicate that CS lines, if
arising from a homogeneous 20$''$ sized region, are optically thin ($\tau$\,(CS 3--2)
$\sim$ 0.086). Our C$^{34}$S and $^{13}$CS spectra thus imply $N$(CS)/$N$(C$^{34}$S)
$\sim$ 12 and $N$(CS)/$N$($^{13}$CS) $\sim$ 50. If, however, clumping within the
assumed 20$''$ sized emission region is substantial and source filling factors
become $f_{\rm c}<$\,0.1, CS line saturation may play a role, leading to higher
$N$(CS)/$N$(C$^{34}$S) and $N$(CS)/$N$($^{13}$CS) ratios. The latter also holds
if our tentative $^{13}$CS profile is an artefact.

\begin{figure}[ht]
\begin{minipage}[t] {18.5cm}
\resizebox{\hsize}{!}{\includegraphics[angle=-90]{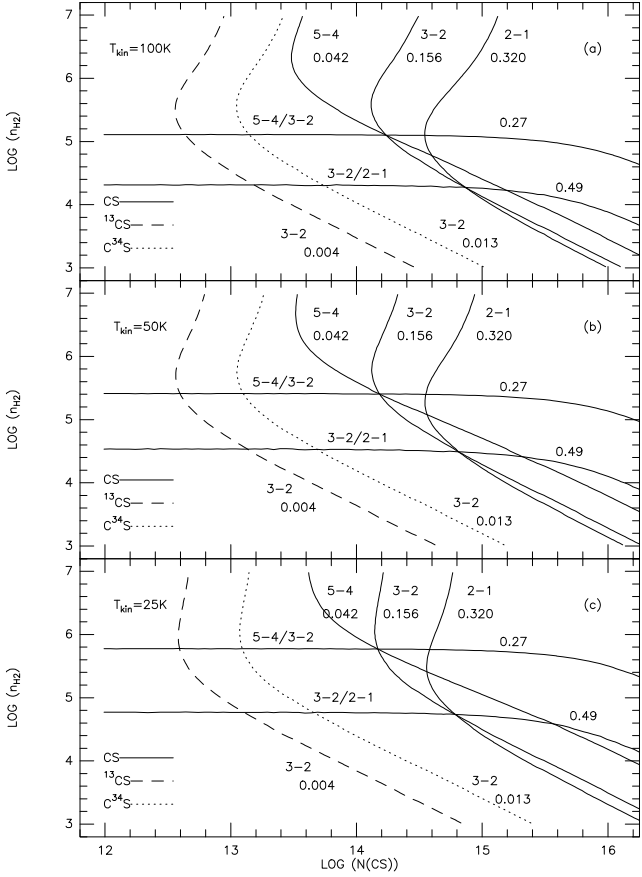}}
\end{minipage}
\caption[]{LVG calculations for CS and its isotopomers C$^{34}$S and $^{13}$CS
at $T_{\rm kin}$\,$\sim$\,100\,K (a), $T_{\rm kin}$\,$\sim$\,50\,K (b) and
$T_{\rm kin}$\,$\sim$\,25\,K (c). }
\end{figure}

\subsection{HC$_3$N}

Cyanoacetylene (HC$_3$N) is another probe of the higher density
gas, providing a particularly large number of transitions that
are, unlike those of CO and possibly CS, almost certainly all
optically thin. We observed 9 rotational HC$_3$N lines, 6 or 7 of
which were detected (first two panels and the first spectrum of
the third panel of Fig.\,4).  Including a total of 22 lines and
assuming that the $J$=17--16 line was detected, the LVG fit (see
Mauersberger et al. 1990) gives for $T_{\rm kin}$=100\,K
$N$(HC$_3$N) $\sim$ 7$\times$10$^{13}$\,cm$^{-2}$ and $n_{\rm
H_2}$ $\sim$ 10$^5$\,cm$^{-3}$ (Fig.\,9), the latter in good
agreement with the results from CS. For smaller kinetic
temperatures, H$_2$ densities would be even larger. The maximum
optical depth is $\tau$\,(HC$_3$N 12-11)\,$\sim$\,0.0024 (100K),
or $\tau$\,(HC$_3$N 10-9)\,$\sim$\,0.0030 (50K). The clumping
factor would therefore have to be small, of order $f_{\rm
c}\sim$\,0.005 or less, to induce line saturation. According to
the model, the $J$=21--20 line is an order of magnitude weaker
than the tentatively detected $J$=17--16 feature. Therefore it is
not surprising that the $J$=24--23 and 25--24 transitions remain
undetected. If the tentative $J$=17--16 line profile is not an
artefact, the 145.6\,GHz feature is the superpostion of two lines
with approximately equal intensities, the $J$=16--15 line of
HC$_3$N and the 2$_{02}$--1$_{01}$ line of ortho-H$_2$CO
(Fig.\,4). Otherwise, H$_2$CO is dominating the line profile.

\begin{figure}[ht]
\begin{minipage}[t]{13.0cm}
\resizebox{\hsize}{!}{\includegraphics[angle=-90]{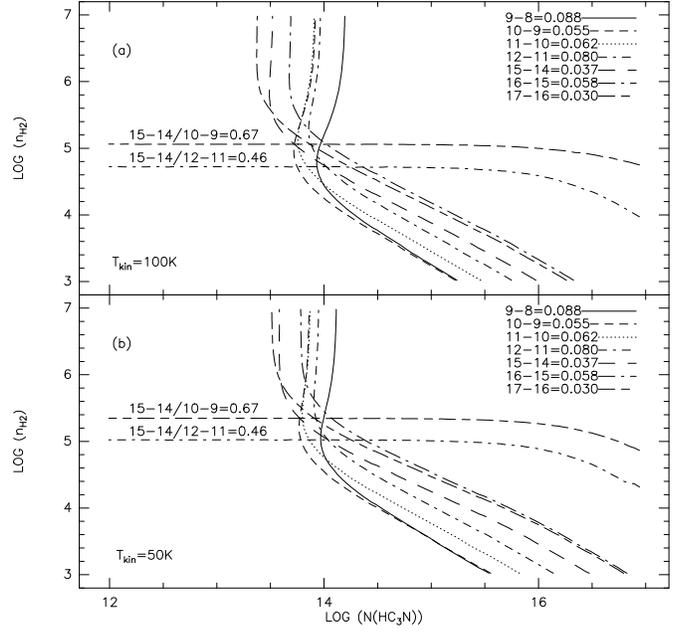}}
\end{minipage}
\caption[]{LVG calculations for HC$_3$N
at $T_{\rm kin}$\,$\sim$\,100\,K (a) and $T_{\rm kin}$\,$\sim$\,50\,K (b).}
\end{figure}

\subsection{SO}

Seven lines of sulfur monoxide (SO) were observed, two of which
were detected (first two panels of Fig.\,5). As a comparison with
the HC$_3$N $J$=10--9 and 12--11 line intensities shows (Fig.\,4),
the 100.1\,GHz profile is dominated by HC$_3$N 11--10, not by SO
4$_5-4_4$ emission. An LVG fit to the SO 3$_2$--2$_1$ and
4$_3$--3$_2$ lines (Fig.\,10; for molecular constants and
collision rates, see Tiemann 1974; Green 1994) gives for $T_{\rm
kin}$=100\,K, $N$(SO)\,$\sim$\,1$\times$10$^{14}$\,cm$^{-2}$ and
$n_{\rm H_2}$\,$\sim$\,10$^5$\,cm$^{-3}$. For $T_{\rm kin}$=50\,K,
$n_{\rm H_2}$ is about 50\% larger while $N$(SO) is not
significantly changed. For the two detected lines source averaged
optical depths are of order 0.005. Densities are high enough to yield
excitation temperatures that are approaching $T_{\rm kin}$. The
computed brightness temperature of the ground state 1$_0$--0$_1$
transition ($\nu \sim$\,30\,GHz) is $\sim$\,35\% of the lines
detected by us.

\begin{figure}[ht]
\begin{minipage}[t]{9.3cm}
\resizebox{\hsize}{!}{\includegraphics[angle=-90]{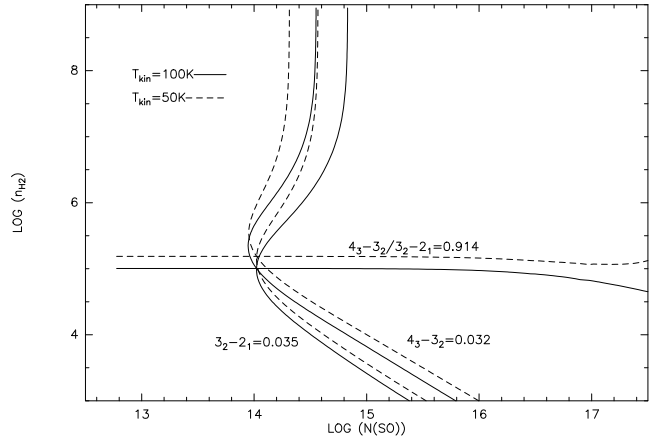}}
\end{minipage}
\caption[]{LVG calculations for SO at $T_{\rm kin}$\,$\sim$\,50\,K
(solid lines) and $T_{\rm kin}$\,$\sim$\,100\,K (dashed lines).}
\end{figure}

\subsection{HCN}

H94 presented an HCN $J$=1--0 map that showed that the emission is
not spatially resolved by the $\sim$\,50$''$ beam of the SEST. C01
detected the HCN $J$=1--0 and 3--2 transitions, while Chin et al.
(1999) reported the detection of the $J$=1--0 lines of HCN,
H$^{13}$CN, and HC$^{15}$N, indicating a surprisingly small
$^{14}$N/$^{15}$N isotope ratio. Here we combine the three
$J$=1--0 transitions with detections of the HCN and H$^{13}$CN
$J$=3--2 lines and a highly tentative detection of the HCN
$J$=4--3 transition (Fig.\,3). The undetected DCN $J$=2--1 line is
bracketed by two comparatively strong lines, those of
CH$_3$OH\,(3--2) and C$^{34}$S\,(3--2), so that even the baseline
remains undetermined (see Fig.\,3).

LVG fits with collision rates from Green \& Thaddeus (1974) are
shown in Fig.\,11. In M\,82, the bulk of the HCN emission arises
from gas that is not only denser but also cooler than the CO
emitting gas (e.g. Mao et al. 2000). We therefore present data
with $T_{\rm kin}$=50 and 100\,K to study differences in the model
parameters. For $T_{\rm kin}$=100\,K and assuming optically thin
emission we obtain a density of $n_{\rm
H_2}$\,$\sim$\,1.5$\times$10$^5$\,cm$^{-3}$ and column densities
of $N$(HCN)\,$\sim$\,2.6$\times$10$^{14}$\,cm$^{-2}$,
$N$(H$^{13}$CN)\,$\sim$\,1.7$\times$10$^{13}$\,cm$^{-2}$ and
$N$(HC$^{15}$N)\,$\sim$\,9.3$\times$10$^{12}$\,cm$^{-2}$. For
$T_{\rm kin}$=50\,K we find instead $n_{\rm
H_2}$\,$\sim$\,3$\times$10$^5$\,cm$^{-3}$. Excitation temperatures
in the low-$J$ lines are $\sim$\,8 and 6\,K for $T_{\rm kin}$=100
and 50\,K, respectively. The lines are subthermally excited.
Therefore the HCN $J$=4--3 line should be 3--4 times weaker than
the $J$=3--2 line, which is consistent with the observations.
Optical depths are of order 0.05--0.1 for the main species. This
implies that line saturation likely affects the HCN $J$=1--0 line
and possibly also the $J$=3--2 transition. Assuming Local
Thermodynamic Equilibrium (LTE) and taking excitation temperatures
from the LVG simulations,
$N$(H$^{13}$CN)$\sim$2.6$\times10^{13}$cm$^{-2}$ from the
intensity of the $J$=1--0 line. With $^{12}$C/$^{13}$C$\sim$50
(Sect.~6.1), $N$(HCN)$\sim$1.3$\times10^{15}$cm$^{-2}$.

Apparently, the gas emitting at 710\,km\,s$^{-1}$ is more highly excited than that observed
at lower velocities. In the $J$=3--2 line the HCN peak is observed at 710\,km\,s$^{-1}$, while
the feature is less dominant in the $J$=1--0 transition (see also Sects.\,5.1.3 and 5.3).

The HCN/H$^{13}$CN $J$=1--0 line intensity ratio of 20--25,
reflecting an abundance ratio of $\sim$\,15 in the optically thin
limit, is similar to that found for the corresponding CO/$^{13}$CO
line intensity ratios. This indicates similar optical depths in
the CO and HCN main isotopomers, while the line emitting volume
may be larger in the case of CO. The line intensity ratios
between the main and the $^{13}$C bearing species can be taken as
lower limits to the $^{12}$C/$^{13}$C isotope ratio.

\begin{figure}[ht]
\begin{minipage}[t]{13.0cm}
\resizebox{\hsize}{!}{\includegraphics[angle=-90]{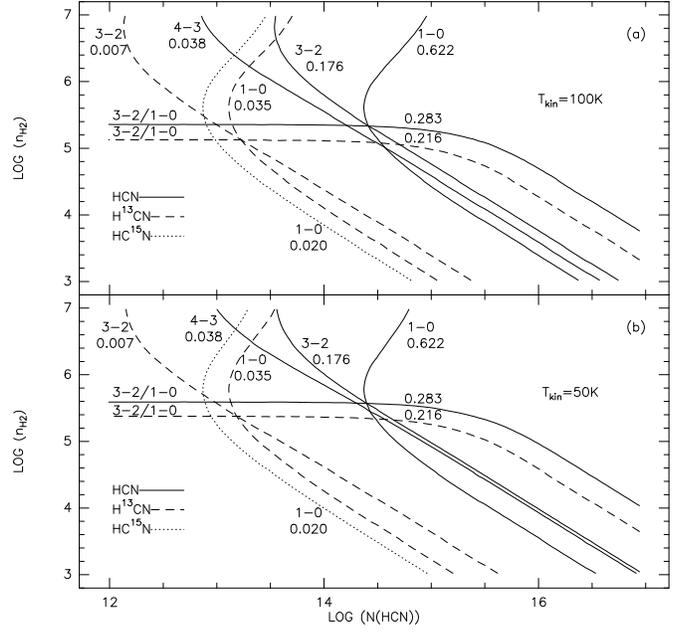}}
\end{minipage}
\caption[]{LVG calculations for HCN (solid lines) and its
isotopomers H$^{13}$CN (dashed lines) and HC$^{15}$N (dotted line)
at $T_{\rm kin}$\,$\sim$\,100\,K (a) and $T_{\rm
kin}$\,$\sim$\,50\,K (b).}
\end{figure}

\subsection{H$_2$CO}

Formaldehyde (H$_2$CO) line profiles are displayed in the last two
panels of Fig.\,2. Fig.\,12 shows results from our LVG
calculations, including the observed $K_a$=1 transitions of
ortho-H$_2$CO, i.e. the 2$_{1,2}$--1$_{1,1}$, 2$_{1,1}$--1$_{1,0}$
and 3$_{1,2}$--2$_{1,1}$ lines (for details of the code, see
Henkel et al. 1980), for $T_{\rm kin}$=100 and 50\,K. For $T_{\rm
kin}$=100\,K we obtain
$N$(ortho-H$_2$CO)\,$\sim$\,10$^{14}$\,cm$^{-2}$ and $n_{\rm
H_2}$\,$\sim$\,4$\times$10$^5$\,cm$^{-3}$. Optical depths are highest in
the $J$=2--1 lines but do not exceed 0.013.  Excitation
temperatures are below 40\,K.

\begin{figure}[ht]
\begin{minipage}[t]{9.3cm}
\resizebox{\hsize}{!}{\includegraphics[angle=-90]{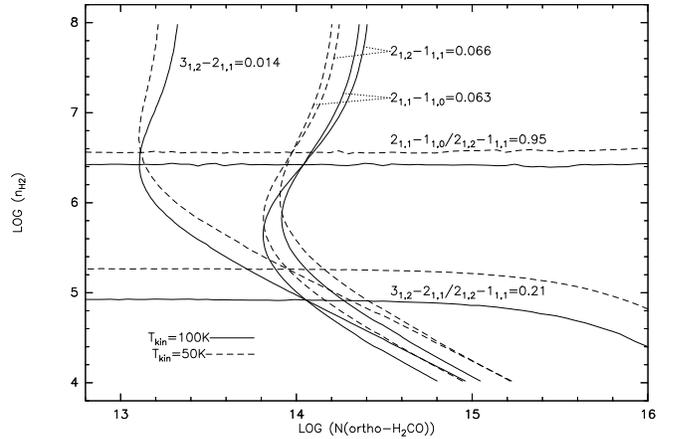}}
\end{minipage}
\caption[]{LVG calculations for ortho-H$_2$CO at $T_{\rm kin}$\,$\sim$\,100\,K
(solid lines) and $T_{\rm kin}$\,$\sim$\,50\,K (dashed lines).}
\end{figure}

\subsection{OCS}

For carbonyl sulfide (OCS) we have a tentative detection of the
$J$=9--8 line (Fig.\,4). If the detection is real, we obtain with
the LVG code already used by Mauersberger et al. (1995) and
assuming $T_{\rm kin}$=100\,K
$N$(OCS)\,$\sim$\,5$\times$10$^{14}$\,cm$^{-2}$ (Fig.\,13). As can
be seen in Fig.\,13, the column density is not significantly
changed for $T_{\rm kin}$=50 and 25\,K, while the H$_2$ density is
poorly constrained.

\begin{figure}[ht]
\begin{minipage}[t]{9.3cm}
\resizebox{\hsize}{!}{\includegraphics[angle=-90]{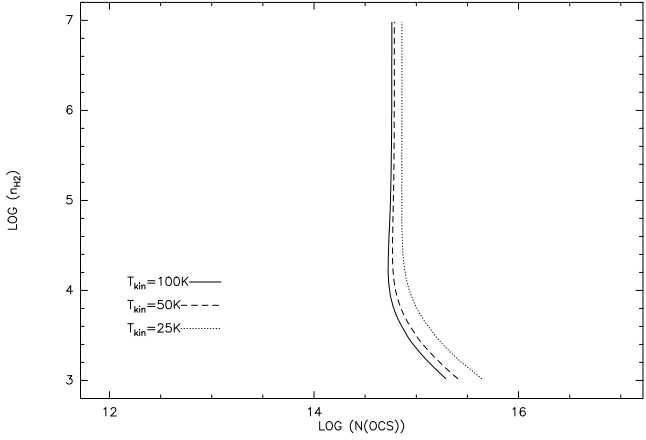}}
\end{minipage}
\caption[]{LVG calculations for OCS at $T_{\rm kin}$\,$\sim$\,100\,K (solid line),
$T_{\rm kin}$\,$\sim$\,50\,K (dashed line), and $T_{\rm kin}$\,$\sim$\,25\,K (dotted line).}
\end{figure}

\subsection{CH$_3$OH}

Eight methanol (CH$_3$OH) features were searched for, three of
which were detected (first two panels of Fig.\,2). The $J$=2--1
profile at 96\,GHz is composed of 4 transitions, and the $J$=3--2
feature at 145\,GHz is composed of 8 lines. An LVG source code for
methanol (CH$_3$OH) was kindly provided by S. Leurini. Einstein
coefficients were taken from Cragg et al. (1993), energy levels
were adopted from Xu \& Lovas (1997), and collision rates are from
Pottage et al. (2001, 2002) for $T_{\rm kin}$\,$\sim$\,20K and
from D. Flower (priv. comm.) for higher $T_{\rm kin}$. Because of
the large linewidths observed towards NGC\,4945, CH$_3$OH lines
are often blended and cannot be observed individually. In our
calculations we therefore added the line intensities of the four
$J$=2--1 transitions near 96\,GHz, of the eight $J$=3--2
transitions near 145\,GHz and of the sixteen $J$=5--4 transitions
near 241\,GHz, respectively. The $J$=2--1 and 3--2 lines are found
within a velocity interval of $\sim$\,80\,km\,s$^{-1}$ (25 and
40\,MHz, respectively) that is small with respect to the linewidth
of the source. The observed $J$=5--4 feature is only slightly
wider than 300\,km\,s$^{-1}$ (Table~2). Thus only few of the 16
lines can contribute significantly to the total intensity.
According to the model, the 5$_{-1}$--4$_{-1}$\,E transition
should dominate. In view of the weak HNCO 10$_{0,10}$--9$_{0,9}$
line (Fig.\,3) we assume that the contribution from the HNCO
11$_{0,11}$--10$_{0,10}$ line is negligible and that the observed
profile is entirely due to CH$_3$OH $J$=5--4 emission. The results
(Fig.\,14) show that for $T_{\rm kin}$=100\,K,
$N$(CH$_3$OH)\,$\sim$\,5.5$\times$10$^{14}$\,cm$^{-2}$ and $n_{\rm
H_2}$\,$\sim$\,1$\times$10$^4$\,cm$^{-3}$. For $T_{\rm
kin}$=25\,K, the column density is similar, but the density
becomes a few times 10$^4$\,cm$^{-3}$. Optical depths are less
than 0.03 for a source size of 20$''$.

\begin{figure}[ht]
\begin{minipage}[t]{13.0cm}
\resizebox{\hsize}{!}{\includegraphics[angle=-90]{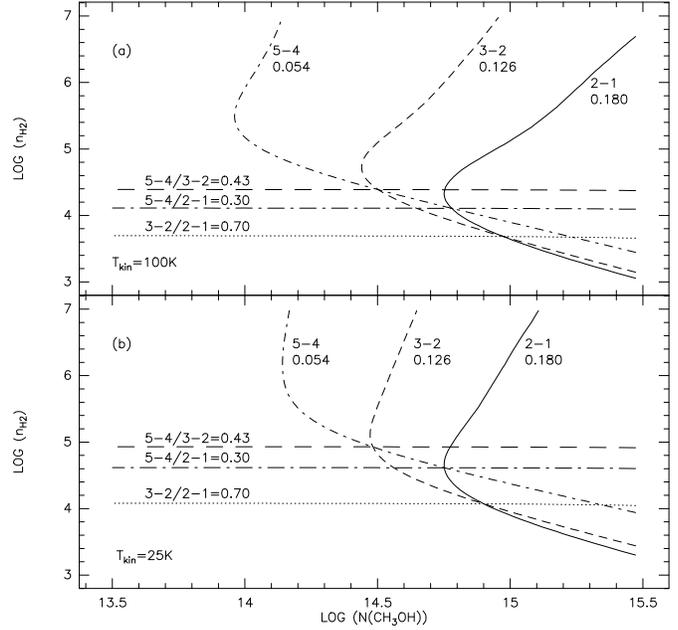}}
\end{minipage}
\caption[]{LVG calculations for CH$_3$OH
with $T_{\rm kin}$\,$\sim$\,100\,K (a) and $T_{\rm kin}$\, $\sim$\,25\,K (b). }
\end{figure}

\subsection{HNCO}

Isocyanic acid (HNCO), observed in six rotational transitions (see
Fig.\,3), is a particularly interesting molecule because it can
probe the far-infrared radiation field ($\lambda\sim$\,330$\mu$m)
with its $K_a$=1 ladder (see Churchwell et al. 1986). As a
consequence, the rotational population distribution of HNCO can be
dominated by far-infrared radiation.

For the radiative transfer calculations of HNCO (see Lapinov et
al. 2001) collision rates were taken from S. Green
(http://www.giss.nasa.gov/data/mcrates/\#hnco). Taking a
background temperature of 2.73\,K and ignoring the
11$_{0,11}-10_{0,10}$ line that is blended by a CH$_3$OH feature
(see Sect.\,4.8), we obtain for $T_{\rm kin}$=100\,K
$N$(HNCO)\,$\sim$\,2.3$\times$10$^{14}$\,cm$^{-2}$ and $n_{\rm
H_2}$\,$\sim$\,1.6$\times$10$^4$\,cm$^{-3}$. Optical depths are
less than 0.02.

Fig.\,3 shows the 4$_{0,4}$--3$_{0,3}$ (or 4$_0$--3$_0$) and 4$_{1,4}$--3$_{1,3}$ (or
4$_{-1}$--3$_{-1}$) line profiles. While the former $K_a$=0 transition is clearly detected,
the latter $K_a$=1 line may be an artefact. The line temperature ratio is therefore $\ga$5:1.
Reproducing a line intensity ratio of 5:1 with the model, we find a background temperature
of $T_{\rm bg}\sim\,$30\,K (see the lower panel of Fig.\,15), that can be taken as an upper
limit to the actual background temperature in the inner 20$''$ of NGC\,4945. Consistent with
Churchwell et al. (1986), our excitation temperatures then reach an equilibrium with the
radiation field so that observed line intensity ratios fail to constrain H$_2$ densities.
The solution with $T_{\rm bg}$=30\,K cannot provide more than a crude approximation. In
contrast to our solution with $T_{\rm bg}$=2.73\,K (Fig.\,15a), lines with measured intensities
do not intersect in Fig.\,15b and are thus not properly reproducing our measurements. A
realistic approach, however, also requires the introduction of a `dilution factor' that
remains observationally unconstrained.

\begin{figure}[ht]
\begin{minipage}[t]{13.0cm}
\resizebox{\hsize}{!}{\includegraphics[angle=-90]{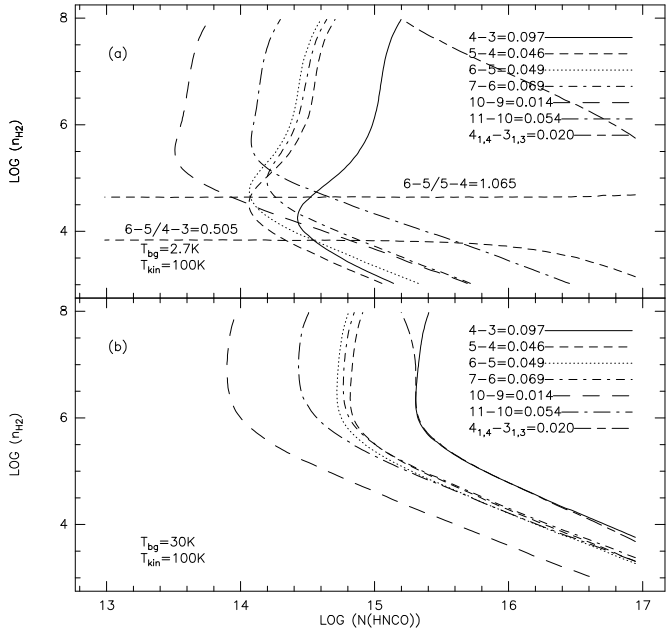}}
\end{minipage}
\caption[]{LVG calculations for HNCO
at $T_{\rm kin}$\,$\sim$\,100\,K and $T_{\rm bg}$\,$\sim$\,2.7\,K (a) and
$T_{\rm bg}$\,$\sim$\,30\,K (b).
The latter case does not allow us to determine H$_2$ densities, so that
horizontal lines (near $\sim$10$^3$\,cm$^{-3}$) are not shown. }
\end{figure}

\begin{table}
\begin{threeparttable}
\begin{scriptsize}
\begin{minipage}{180mm}
\caption[]{Intensities used for LVG and LTE calculations}
\begin{flushleft}
\begin{tabular}{lrl}
\hline
Transition & $T^{ \prime}_{\rm mb}$  & $\eta_{\rm bf}$\tnote{a)}  \\
& (K) & \\
\hline\hline
CO 1--0                       &6.143 &0.293 \\
CO 2--1                       &7.080 &0.452 \\
CO 3--2                       &7.000 &0.500 \\
$^{13}$CO 1--0                &0.410 &0.293 \\
$^{13}$CO 2--1                &0.531 &0.452 \\
$^{13}$CO 3--2                &0.600 &0.500 \\
C$^{18}$O 1--0                &0.102 &0.293 \\
C$^{18}$O 2--1                &0.155 &0.452 \\
C$^{17}$O 1--0                &0.014 &0.293 \\
C$^{17}$O 2--1                &0.027 &0.452 \\
\hline\hline
CS 2--1                       &0.320 &0.125 \\
CS 3--2                       &0.156 &0.243 \\
CS 5--4                       &0.042 &0.471 \\
C$^{34}$S 3--2                &0.013 &0.237 \\
$^{13}$CS 3--2                &0.004 &0.222 \\
\hline\hline
CH$_3$OH 2--1                 &0.180 &0.122 \\
CH$_3$OH 3--2                 &0.126 &0.238 \\
CH$_3$OH 5--4                 &0.054 &0.465 \\
\hline\hline
H$_2$CO 2$_{1,2}$--1$_{1,1}$  &0.066 &0.228 \\
H$_2$CO 2$_{1,1}$--1$_{1,0}$  &0.063 &0.252 \\
H$_2$CO 3$_{1,2}$--2$_{1,1}$  &0.014 &0.431 \\
\hline\hline
HNCO 4$_{0,4}$--3$_{0,3}$       &0.097 &0.103 \\
HNCO 5$_{0,5}$--4$_{0,4}$       &0.046 &0.152 \\
HNCO 6$_{0,6}$--5$_{0,5}$       &0.049 &0.205 \\
HNCO 7$_{0,7}$--6$_{0,6}$       &0.069 &0.260 \\
HNCO 10$_{0,10}$--9$_{0,9}$       &0.014 &0.418 \\
HNCO 11$_{0,11}$--10$_{0,10}$       &$<$0.054 &0.465 \\
\hline\hline
HCN 1--0                        &0.622 &0.105 \\
HCN 3--2                        &0.176 &0.512 \\
HCN 4--3                        &0.038 &0.651 \\
H$^{13}$CN 1--0                 &0.035 &0.100 \\
H$^{13}$CN 3--2                 &0.007 &0.499 \\
HC$^{15}$N 1--0                 &0.020 &0.099 \\
\hline\hline
HC$_3$N 9--8                    &0.088 &0.091 \\
HC$_3$N 10--9                   &0.055 &0.110 \\
HC$_3$N 11--10                  &0.062 &0.130 \\
HC$_3$N 12--11                  &0.080 &0.151 \\
HC$_3$N 15--14                  &0.037 &0.217 \\
HC$_3$N 16--15                  &0.058 &0.240 \\
HC$_3$N 17--16                  &0.030 &0.262 \\
\hline\hline
SO 3$_2$--2$_1$                 &0.035 &0.128 \\
SO 4$_3$--3$_2$                 &0.032 &0.221 \\
\hline\hline
OCS 9--8                        &0.033 &0.151 \\
\hline\hline
CN 1-0 $J$=3/2--1/2             &0.436 &0.161 \\
CN 1-0 $J$=1/2--1/2             &0.263 &0.160 \\
CN 2-1 $J$=5/2--3/2             &0.162 &0.433 \\
\hline\hline

HNC 1--0                        &0.413 &0.109 \\
HN$^{13}$C 1--0                 &0.028 &0.101 \\
\hline\hline
HCO$^+$ 1--0                    &0.797 &0.106 \\
HCO$^+$ 3--2                    &0.115 &0.520 \\
\hline
\end{tabular}
\begin{tablenotes}
\item[a)] for the definition of $\eta_{\rm bf}$, see Sect.\,4
\end{tablenotes}
\end{flushleft}
\end{minipage}
\end{scriptsize}
\end{threeparttable}
\end{table}

\section{Other molecular species}

In Sect.\,4 a variety of molecular species were analyzed with an
LVG code. Other molecules, mostly with a smaller number of
detected transitions, were analyzed with an LTE approach. In the
following some of these molecules are introduced, among them CN
with the $N$=1--0 and 2--1 lines being observed.

\subsection{CN}

\subsubsection{The [CN]/[$^{13}$CN] abundance ratio}

For the CN $N$=1--0 transition we find a group of $J$=1/2--1/2 lines ($\nu\sim$\,113.17\,GHz)
and a group of $J$=3/2--1/2 lines ($\nu\sim$\,113.5\,GHz) that were detected with high
signal-to-noise ratios (Fig.\,4). According to Henkel et al. (1998), CN provides the most
stringent limits on $^{12}$C/$^{13}$C isotope ratios in extragalactic molecular clouds.
Because the $N$=1--0 transition is split into several lines (spin-rotation and nuclear
magnetic and electric quadrupole interactions), optical depths are not large and can be
determined by an analysis of line intensity ratios. Deviations from LTE (Local Thermodynamic
Equilibrium) intensities appear not to be large.

In order to determine the optical depth of the CN $N$=1--0 lines, we denote the $J$=3/2--1/2
lines with the index 1 and the $J$=1/2--1/2 lines with the index 2. With an observed line
intensity ratio of 1.58$\pm$0.06 (Table~2) and $\tau_1$=2$\tau_2$ (see Skatrud et al. 1983
for relative LTE intensities under optically thin conditions) we then find
$$
\frac{\int T_{\rm mb,1} \mathrm{d}v}{\int T_{\rm mb,2} \mathrm{d}v}
=\frac{1-\mathrm{e}^{-\tau_1}}{1-\mathrm{e}^{-\tau_2}}
=\frac{1-\mathrm{e}^{-\tau_1}}{1-\mathrm{e}^{-\tau_1/2}}=1.58\pm0.06.
$$
For $\tau_1$ we obtain $\tau_1$=1.09$\pm$0.21

The $J$=3/2--1/2 and 1/2--1/2 lines are part of the same spectrum so that their line intensity
ratio is not affected by calibration errors. With $\tau_1$ and $\tau_2$ known and
$$
I_1 \propto \frac{1}{\Sigma g_{i,1}}\,\cdotp\,\int T_{\rm mb,1} \mathrm{d}v\,\cdotp\,\frac{\tau_1}{1-\mathrm{e}^{-\tau_1}}
$$
and
$$
I_2 \propto \frac{1}{\Sigma g_{i,2}}\,\cdotp\,\int T_{\rm mb,2} \mathrm{d}v\,\cdotp\,\frac{\tau_2}{1-\mathrm{e}^{-\tau_2}},
$$
where $\Sigma g_{i,1}$ and $\Sigma g_{i,2}$ are 0.667 and 0.333, respectively (Skatrud et al.
1983), we then obtain for the `optical depth corrected' total CN $N$=1--0 line intensities
$I_1$=$I_2$=60.4$\pm$5.4\,K\,km\,s$^{-1}$.

For $^{13}$CN (spectra are shown in Fig.\,4) we took the frequencies and relative line
intensities (LTE, $\tau$$\ll$1) from Bogey et al. (1984) and Gerin et al. (1984). With a
3$\sigma$ noise level of 5.7\,mK in an 18.82\,km\,s$^{-1}$ wide channel (see Table~2 and
Fig.\,4) we obtain for a 300\,km\,s$^{-1}$ wide line an upper limit to the integrated
intensity of
$$
3\sigma_{\int T dv}=5.7\,\mathrm{mK\,\cdotp\,18.82\,km\,s^{-1}}\,\cdotp\,\sqrt{300/18.82}
$$
The resulting 3$\sigma$ upper limit is 0.43\,K\,km\,s$^{-1}$.

As in the case of CN we have to correct for the fact that the $^{13}$CN $N$=1--0 lines are
distributed over a frequency range (108.4--108.8\,GHz). In contrast to CN, $^{13}$CN does
not require any corrections related to line saturation. The most prominent group of $^{13}$CN
lines, located between 108.7802 and 108.7964\,GHz (the difference corresponds to $\sim$\,45\,km\,s$^{-1}$
that is small with respect to a total linewidth of 300\,km\,s$^{-1}$) comprises 42.7\% of the
total LTE intensity. As a consequence we find
$$
[^{12}\mathrm{CN}]/[^{13}\mathrm{CN}]\geq\,60.4/(0.43/0.427)\,\sim\,60.
$$
Henkel et al. (H94) reported a 3$\sigma$ lower limit to the HCN/H$^{13}$CN line intensity
ratio of 31, while the actual ratio is 23.4$\pm$1.9 (Table~2). In view of this discrepancy
we conservatively estimate that the [$^{12}$CN]/[$^{13}$CN] abundance ratio should be $\sim$45
or larger.

\subsubsection{CN line intensity anomalies}
\ \\
The $N$=1--0 transition: \\

As already mentioned, the CN transitions represent the
superposition of a number of hyperfine (HF) components that are
displayed in Fig.\,16. While the $N$=1--0 $J$=3/2--1/2 profile
(`HF1' in Fig.\,16a) consists of a broad blue- and a narrow
redshifted line, the $J$=1/2--1/2 profile (`HF2' in Fig.\,16a) is
wider and shows a single broad feature (see also the observed
spectra in Fig.\,4). Is this difference caused by a different
intrinsic lineshape? A fit with two velocity components,
characterized by $v_1$=450\,km\,s$^{-1}$, $\Delta
v_{1/2,1}$=300\,km\,s$^{-1}$ and $v_2$=710\,km\,s$^{-1}$, $\Delta
v_{1/2,2}$=50\,km\,s$^{-1}$ reproduces the observed $J$=3/2--1/2
and 1/2--1/2 profiles in a satisfactory way (the linewidths refer
to full width to half power Gaussians in optical depth). The lack
of a prominent redshifted spike in the $J$=1/2--1/2 lines is not
caused by a different intrinsic lineshape but is a consequence of
the different frequencies and strengths of the individual HF
components (see Fig.\,16).

For the fit to the line profile we assumed optical depth ratios between individual HF components
that are consistent with the relative intensities given by Skatrud et al. (1983). To scale
optical depths, we took $\Sigma\tau_i$=1.09 for the $J$=3/2--1/2 and $\Sigma\tau_i$=0.545
($\tau_1$=2$\tau_2$, see Sect.\,5.1.1) for the $J$=1/2--1/2 HF components, the index $i$ denoting
individual HF features. It is remarkable that the sum of these optical depths is, within the
limits of noise, the same for the blue- ($\sim$450\,km\,s$^{-1}$) and redshifted
($\sim$710\,km\,s$^{-1}$) velocity components. \\

\begin{figure}[ht]
\resizebox{\hsize}{!}{\includegraphics[angle=0]{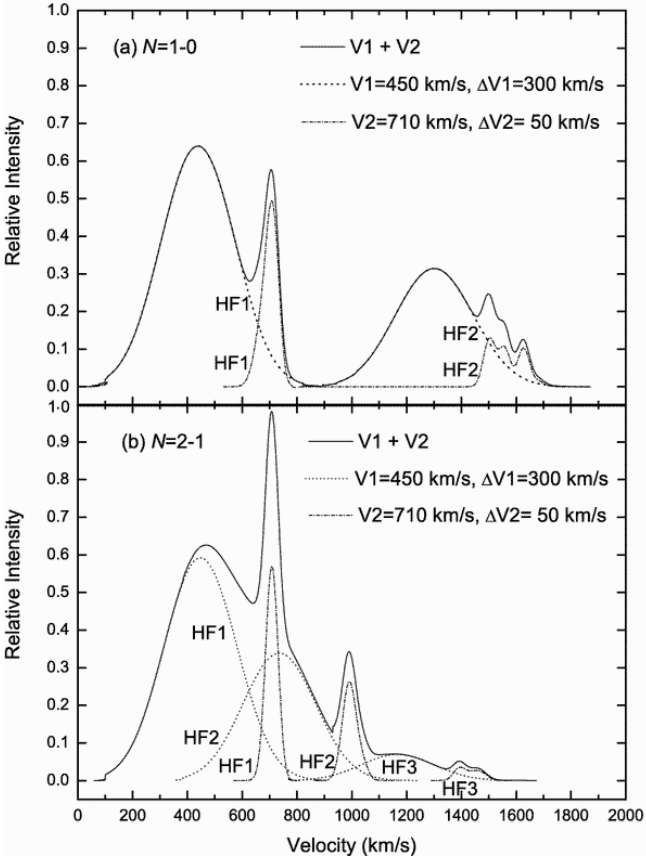}}
\caption[]{a) Upper panel: The computed CN $N$=1--0 profile
composed of the two velocity components V1 and V2 (see also
Sect.\,5.1.2). Observed CN profiles are displayed in Fig.\,4. It
is apparent that the strong spike in the $J$=3/2--1/2 line (HF1)
at 710\,km\,s$^{-1}$ has no similarly prominent counterpart in the
lower frequency $J$=1/2--1/2 line (HF2), located at
$\sim$1500--1600\,km\,s$^{-1}$ in the rest frequency frame of the
3/2--1/2 (HF1) features. b) Lower panel: The CN $N$=2--1 spectrum
computed with the same velocity components V1 and V2 that were
used to simulate the 1--0 spectrum in the upper panel. For the
$N$=2--1 transition, three groups of hyperfine components (HF1,
HF2 and HF3) had to be taken into account. Note that the computed
intensity ratio of the peaks at 710 and 450\,km\,s$^{-1}$ is
$\sim$1.5, while the observed spectrum (Fig.\,4) shows a ratio
$\sim$1.9.}
\end{figure}

\ \\
The $N$=2--1 transition: \\

A comparison of the $N$=1--0 $J$=3/2--1/2 and $N$=2--1 CN profiles
was given by H94, who suggested that the relative strength of the
$N$=2--1 710\,km\,s$^{-1}$ feature (see Fig.\,4) is higher than
that implied by the $N$=1--0 $J$=3/2--1/2 lineshape. Such an
excess in the $N$=2--1 710\,km\,s$^{-1}$ feature is also indicated
by our new data, making use of the V1 and V2 components defined by
the $N$=1--0 line and subdividing the HF components into three
groups, HF1, HF2, and HF3 (see Fig.16b). The excess becomes
apparent when comparing the observed line intensity ratio between
the 710 and 450\,km\,s$^{-1}$ components, $\sim$1.9 (Fig.\,4),
with that of our computation, $\sim$1.5 (Fig.\,16).

Concluding that the strongest CN $N$=2--1 component, the 710\,km\,s$^{-1}$ feature,
is enhanced, we find that line saturation cannot play a significant role in this
transition. Otherwise, the feature would be weaker, not stronger than expected in our
analysis assuming optically thin emission.

\subsubsection{CN column density}

A comparison of total line intensities gives $\sim$40\,K\,km\,s$^{-1}$ for the CN $N$=1--0
and $\sim$30\,K\,km\,s$^{-1}$ for the $N$=2--1 line (see Table~2). Accounting for the optical
depth in the $N$=1--0 transition, the integral would be $\sim$60\,K\,km\,s$^{-1}$ if we
correct for saturation effects (Sect.\,5.1.1), while line saturation may not be relevant
for the $N$=2--1 line (Sect.\,5.1.2). Correcting these integrated line intensities for beam
size effects, i.e., dividing the integrated line intensities by $\eta_{\rm bf}$ (see Sect.\,4
and Table~4), we obtain $\sim$375\,K\,km\,s$^{-1}$ for the $N$=1--0 line accounting for line
saturation and $\sim$70\,K\,km\,s$^{-1}$ for the $N$=2--1 line. The line temperature ratio is
2--1/1--0$\sim$\,0.185. With
$$
0.185=4\times\,\mathrm{e}^{-x}\times\frac{1-\mathrm{e}^{-2x}}{1-\mathrm{e}^{-x}}\times
\frac{(\mathrm{e}^{2x}-1)^{-1}-(\mathrm{e}^{2y}-1)^{-1}}{(\mathrm{e}^x-1)^{-1}-(\mathrm{e}^y-1)^{-1}}
$$
$x$=h$\nu_{10}$/${\rm k}T_{ex}$, $\nu_{10}$=113.386\,GHz and
$y$=h$\nu_{10}$/${\rm k}\cdot$2.73=1.99 we then obtain for the
excitation temperature $T_{\rm ex}\sim$3.1\,K and for the column
density, assuming LTE, $N$(CN)$\sim$2.5$\times10^{14}$cm$^{-2}$.
We note that the excitation temperature is very small, indicating
densities well below 10$^{5}$\,cm$^{-3}$ (see e.g. Fuente et al.
1995). Most of the $N$=1--0 710\,km\,s$^{-1}$ feature is
originating from the HF1 group of HF components (see Fig.\,2 and
Sect.\,5.1.2), i.e. from gas that is characterized by a
recessional velocity of 710\,km\,s$^{-1}$. If this component has
an $N$=2--1/1--0 ratio that is twice the average value (see H94),
this implies $T_{ex}\sim$4.1\,K (see also Sects.\,4.5 and 5.3).

\subsection{HNC}

The $J$=1--0 HNC and HN$^{13}$C lines (Fig.\,3) were detected with
an intensity ratio of 11.4$\pm$1.4, i.e. well below the
$^{12}$C/$^{13}$C lower limit derived by CN (Sect.\,5.1.1). This
implies that the HNC $J$=1--0 line must be optically thick. With a
smaller $^{12}$C/$^{13}$C line intensity ratio than measured for
HCN (Sect.\,4.5) and with only half the integrated intensity, HNC
is likely arising from a more compact region than HCN and HCO$^+$.
Assuming LTE and an excitation temperature of $T_{\rm
rot}\sim$8\,K that was derived for HCN, a molecule with similar
dipole moment and rotational constant, we obtain
$N$(HN$^{13}$C)$\sim$3.5$\times10^{13}$cm$^{-2}$. With
$^{12}$C/$^{13}$C$\sim$50 (Sect.\,6.1),
$N$(HNC)$\sim$1.8$\times10^{15}$cm$^{-2}$.

\subsection{HCO$^+$, HCO and thermal SiO}

While the $J$=1--0 profile of HCO$^+$ shows, as other lines, a dominant component at 450\,km\,s$^{-1}$
and a satellite component at 710\,km\,s$^{-1}$, the relative strength of the features is inverted in
the $J$=3--2 spectrum ({see Fig.\,5}). Apparently, the gas seen at 710\,km\,s$^{-1}$ is higher excited
(see also Sects.\,4.5 and 5.1.3).

H$^{13}$CO$^+$ is blended with the HCO 1$_{01}-0_{00}$ and SiO $J$=2--1 $v$=0 transitions (see
Garc\'{\i}a-Burillo et al. 2002). In order to provide meaningful constraints on line intensity ratios,
the three profiles shown in Fig.\,5 (lower right corner) were simulated. Assuming that the 450 and
710\,km\,s$^{-1}$ components of a given molecular transition have equal intensities, and linewidths of
200 and 150\,km\,s$^{-1}$, respectively, we find that SiO might be strongest while HCO might be weakest.
With $I_{\mathrm{HCO}(F=1-0)}$/$I_{\mathrm{HCO}(F=2-1)}\,\sim$\,0.3, for the HCO 1$_{01}-0_{00}$
$J$=3/2--1/2 lines we find qualitatively $I_{\mathrm{H}^{13}\mathrm{CO}^+}$/$I_{\mathrm{HCO}(F=2-1)}\,\sim$\,2.0
and $I_{\mathrm{SiO}(2-1 v=0)}$/$I_{\mathrm{HCO}(F=2-1)}\,\sim$\,2.5.

The resulting HCO$^+$/H$^{13}$CO$^+$ line intensity ratio is 41
and excitation temperatures become 4.7\,K (450\,km\,s$^{-1}$) and
5.4\,K (710\,km\,s$^{-1}$ component) in the case of optically thin
emission. We note that the decomposition of SiO, HCO and
H$^{13}$CO$^+$ is too tentative to provide meaningful error
estimates. A search for SiO $J$=3--2 $v$=0 emission that is less
contaminated by other spectral features would be worthwhile. The
HCO$^+$/H$^{13}$CO$^+$ line intensity ratio, being much more
uncertain than other such ratios, is not included in Table~5.

\subsection{C$_3$H$_2$}

Profiles of cyclopropenylidene (C$_3$H$_2$) are shown in the right
panel of Fig.\,2. The 85\,GHz 2$_{12}-1_{01}$ line (Fig.\,2) shows
the double peak structure with one main component at 450 and the
other at 710\,km\,s$^{-1}$ that also characterizes a few other species,
most notably CN (see Fig.\,4 and Sect.\,5.1). Higher excited C$_3$H$_2$
transitions (Fig.\,2), the 86\,GHz 4$_{32}-4_{23}$ line and the
superposition of the 150\,GHz 4$_{04}-3_{13}$ and 4$_{14}-3_{03}$
lines (separation:$\sim$60\,km\,s$^{-1}$) show, however, an
approximately Gaussian lineshape. With the partition function
(e.g. http://spec.jpl.nasa.gov), the electric dipole moment,
asymmetric rotor line strengths and excitation above the ground
state (e.g. Thaddeus et al. 1985; Vrtilek et al. 1987; Lovas \&
Suenram 1989) known, a column density assuming LTE and optically
thin emission can be derived.

Using the 2$_{1,2}-1_{1,0}$ and 4$_{3,2}-4_{2,3}$ lines that both
belong to the ortho-C$_3$H$_2$ species and that were observed with
similar beam sizes, an excitation temperature of 23$\pm$5\,K is obtained.
The total column density for a source size of 20$''$ and an ortho-para
ratio of 3:1 is estimated to be $\sim$3.3$\times10^{14}$\,cm$^{-2}$.

\subsection{C$_2$H}

While Table~2 and Fig.\,5 show the parameters of a one component Gaussian fit to the 87\,GHz
$N$=1--0 transition of C$_2$H, the line is actually split into two components with an intensity
ratio of $\sim$3:1. This is caused by hyperfine splitting (e.g. Tucker et al. 1974; Ziurys et
al. 1982). Two groups of HF components are present, a redshifted one at $\sim$87.32\,GHz and a
blueshifted one at $\sim$87.41\,GHz. Their LTE intensity ratio in the optically thin limit is
2:1. Our observed ratio is higher, likely not because of maser amplification of the stronger
feature, but because some part of the weaker blueshifted HF component, that near 710\,km\,s$^{-1}$,
is superposed onto the dominant more redshifted HF component observed. A simulation of the entire
profile as for CN (see Sect.\,5.1.2) or for the blend of HCO$^+$, HCO and SiO features (Sect.\,5.3),
assuming an intensity ratio of 2:1 for the group of HF components (Tucker et al. 1974; Ziurys et al.
1982), an intensity ratio of 1:1 between the 450 and 710\,km\,s$^{-1}$ features and linewidths of
200 and 150\,km\,s$^{-1}$, respectively, provides a good fit. Our assumption that the emission is
optically thin then yields, with $T_{\rm ex}$=20\,K, $N_{\rm C_2H}\sim6.8\times10^{15}$cm$^{-2}$.

\subsection{CH$_3$CCH and CH$_3$CN}

Extragalactic CH$_3$CCH and CH$_3$CN was first reported by Mauersberger et al. (1991). While
CH$_3$CCH was detected in NGC\,253 and M\,82, CH$_3$CN was only seen in NGC\,253.

In the case of NGC\,4945 we also obtain a clear detection of
CH$_3$CCH, while CH$_3$CN is not detected convincingly (Fig.\,5).
This may only be due to blending by the CS $J$=3--2 line, i.e. a
search for the $J$=5--4, 6--5 or 7--6 transitions of CH$_3$CN
would be meaningful. Applying
$N$(CH$_3$CCH)$\sim2.8\times10^{12}$\,cm$^{-2}$(K\,km\,s$^{-1}$)$^{-1}\cdot
I_{\rm CH_3CCH}$ (Mauersberger et al. 1991), we obtain a column
density of 2.4$\times10^{14}$\,cm$^{-2}$.

\section{Isotope ratios}

\subsection{$^{12}$C/$^{13}$C and $^{16}$O/$^{18}$O ratios}

For a complete evaluation of the column densities discussed in
Sects.\,4 and 5, a knowledge of isotope ratios in the nuclear
region of NGC\,4945 is indispensable. Line intensity ratios of
$^{12}$C and $^{13}$C bearing isotopomers provide direct
information on line saturation in the more abundant species if the
$^{12}$C/$^{13}$C ratio is known. Table~5 summarizes the observed
$^{12}$C/$^{13}$C, $^{14}$N/$^{15}$N, $^{16}$O/$^{18}$O,
$^{16}$O/$^{17}$O and $^{32}$S/$^{34}$S values based on integrated
intensities. To interpret the ratios we assume in the following
that they reflect the isotopic ratios in the case that the lines
are optically thin or that they provide lower limits to the
isotope ratios, if the line of the more abundant species is
optically thick. Fractionation and isotope selective
photodissociation may not be important in the nuclear region of
NGC\,4945 (H94) and are thus ignored. The data do not yet permit a
separate analysis of the 450 and 710\,km\,s$^{-1}$ velocity
components and are therefore based on peak temperatures and
integrated intensities of the entire individual spectra.

There is a total of 8 measures of the $^{12}$C/$^{13}$C ratio (Table~5). Four of them yield
small values (10--15), two intermediate values (15--30) and two high values ($\ga$45). The
comparison of line intensity ratios suggests that CO and HNC rotational lines have highest optical
depths, followed by those of HCN. While it may appear strange, at first sight, that
$\tau$(CO $J$=2--1)$\sim\tau$(HNC $J$=1--0), HNC is presumably arising from a smaller area,
where average H$_2$ column densities are larger than those in the region being responsible for
the bulk of the CO emission. Note that CO/$^{13}$CO line intensity ratios decrease with increasing
rotational quantum number $J$. This is expected in a warm molecular medium where CO column densities
in individual states rise almost proportional to their statistical weights (2$J+$1).

With ratios or lower limits of 40--50, the $^{12}$C/$^{13}$C values derived from CN and CS are
closer to the actual $^{12}$C/$^{13}$C isotope ratio. While CN did allow us to estimate optical
depths, CS may show a direct albeit low signal-to-noise detection of its $^{13}$C bearing species.

\begin{table}
\begin{threeparttable}
\begin{scriptsize}
\begin{minipage}{180mm}
\caption[]{Intensity ratios of isotopomers}
\begin{flushleft}
\begin{tabular}{lcr@{$\pm$}lr@{$\pm$}l}
\hline
Molecules & Transition & \multicolumn{2}{c}{Ratio1\tnote{a)}} & \multicolumn{2}{c}{Ratio2\tnote{b)}}  \\
\hline\hline
CO/$^{13}$CO          & 1--0             & 15.2 & 0.5 &15.0 &2.1\\
                      & 2--1             & 11.3 & 0.1 &13.3 & 1.9\\
                      & 3--2             &  9.9 & 1.1 &11.7 &1.6\\
CN/$^{13}$CN\tnote{d)}&             & \multicolumn{2}{c}{$>$45}   &\multicolumn{2}{c}{...} \\
CS/$^{13}$CS\tnote{c)}& 3--2             & 49.0 & 17.1 &39.0 & 5.5\\
HCN/H$^{13}$CN        & 1--0             & 23.4 & 1.9 &17.8 &2.5\\
                      & 3--2             & 30.0 & 5.2 &25.1 &3.6\\
HNC/HN$^{13}$C        & 1--0             & 11.4 &1.4  &14.8 &2.1 \\
HCN/HC$^{15}$N        & 1--0             &48.3  & 5.1 &31.1 &4.4\\
CO/C$^{18}$O          & 1--0             & 56.3 & 2.5 &60.2 &8.5\\
                      & 2--1             & 37.5 & 0.8 &45.7 &6.5\\
CO/C$^{17}$O          & 1--0             &433.0 &83.2 &438.8&62.1\\
                      & 2--1             &278.3 &26.5 &262.2&37.1\\
$^{18}$CO/C$^{17}$O   & 1--0             &  7.7 & 1.5 &  7.3& 1.0\\
                      & 2--1             &  7.4 & 0.7 &  5.7& 0.8\\
CS/C$^{34}$S          & 3--2             & 10.4 & 0.8 &12.0 &1.7\\

\hline
\end{tabular}
\begin{tablenotes}
\item[a)] Ratios of integrated intensities (see Table~2). Corrections for beam size effects
are less than 10\%.
\item[b)] Ratios of corrected peak intensities divided by $\eta_{\rm bf}$ (see Sect.\,4 and
Table~4)
\item[c)] If $^{13}$CS or H$^{13}$CO$^+$  is not detected, the ratios are lower limits. For
HCO$^+$ only the 710\,km\,s$^{-1}$ component was taken to derive the ratios.

\item[d)] See Sect.\,5.1.1
\end{tablenotes}
\end{flushleft}
\end{minipage}
\end{scriptsize}
\end{threeparttable}
\end{table}

Following H94, we obtain for the CO $J$=1--0 lines
$$
^{12}{\rm C}/^{13}{\rm C} \leq \left[I({\rm CO})/I(^{13}{\rm CO})\right] \,\tau ({\rm CO})
\sim (15\pm2)\,\tau ({\rm CO})
$$
and
$$
^{16}{\rm O}/^{18}{\rm O} \leq \left[I({\rm CO})/I({\rm C}^{18}{\rm O})\right] \,\tau ({\rm CO})
\sim (58\pm9)\,\tau ({\rm CO}),
$$
where the inequality on the left part of both equations accounts for the fact that CO, the main
isotopomer, is optically thick.

For the $J$=2--1 lines we obtain correspondingly
$$
^{12}{\rm C}/^{13}{\rm C} \leq \left[I({\rm CO})/I(^{13}{\rm CO})\right] \,\tau ({\rm CO}) \sim
(12\pm2)\,\tau ({\rm CO})
$$
and
$$
^{16}{\rm O}/^{18}{\rm O} \leq \left[I({\rm CO})/I({\rm C}^{18}{\rm O})\right] \,\tau ({\rm CO})
\sim (41\pm7)\,\tau ({\rm CO}).
$$

If both $^{13}$CO and C$^{18}$O were optically thin, we could then derive, neglecting fractionation
effects and isotope selective photoionization (see H94),
$^{12}{\rm C}/^{13}{\rm C}\sim(0.26\pm0.05)\times(^{16}{\rm O}/^{18}{\rm O})$ for the $J$=1--0 lines
and $^{12}{\rm C}/^{13}{\rm C}\sim(0.29\pm0.07)\times(^{16}{\rm O}/^{18}{\rm O})$ for the $J$=2--1
lines. If, however, also $^{13}$CO is optically thick, the values would be lower and would be closer
to $^{12}{\rm C}/^{13}{\rm C}\sim0.10\times(^{16}{\rm O}/^{18}{\rm O})$ that is found in the nuclear
region of the Milky Way.

As already mentioned, low $J$ CO column densities almost rise with their statistical weight, 2$J+$1,
in a warm approximately thermalized medium. As a consequence,
$\tau$($^{13}$CO $J$=2--1)$>\tau$($^{13}$CO $J$=1--0). The fact that the correlation between
$^{12}$C/$^{13}$C and $^{16}$O/$^{18}$O ratios is almost independent of the choice of transition
($J$=1--0 or $J$=2--1) does not prove but strongly suggests that the $^{13}$CO opacities are small
(see also Bergman et al. 1992 and H94). This implies that the ratio between $^{12}$C/$^{13}$C and
$^{16}$O/$^{18}$O ratios does indeed drastically differ from that in the central region of the Galaxy.
Giving a higher weight to the result from the $J$=1--0 lines because of a lower degree of $^{13}$CO line
saturation and smaller errors, we then find $^{12}$C/$^{13}$C$\ga$40 and $^{16}$O/$^{18}$O$\ga$150.

Inflowing gas through a bar (Ott et al. 2001) might provide fresh material with $^{12}$C/$^{13}$C in
excess of 50, if we use the analogy to the Galaxy with a $^{12}$C/$^{13}$C isotope ratio of $\sim$70
near the solar circle (e.g. Wilson \& Rood 1994). However, in the middle or at the end of a violent
starburst (see the conflicting statements of Koornneef 1993 and Ott et al. 2001 on future star formation
in NGC\,4945), $^{18}$O production might be even more effective than $^{16}$O production (i.e.
$^{16}$O/$^{18}$O ratios may decrease) if metallicities are solar or higher (Henkel \& Mauersberger 1993;
Henkel et al 1994b; Langer \& Henkel 1995). With a $^{16}$O/$^{18}$O ratio of $\sim$250 in the central
region of the Milky Way we then find $^{12}$C/$^{13}$C$\sim$40--60 and $^{16}$O/$^{18}$O$\sim$150--240
for NGC\,4945. A detection of $^{13}$CN in NGC\,4945 could strengthen the case for our upper limits to
the $^{12}$C/$^{13}$C and $^{16}$O/$^{18}$O ratios.

\subsection{$^{18}$O/$^{17}$O, $^{14}$N/$^{15}$N and $^{32}$S/$^{34}$S ratios}

An $^{18}$O/$^{17}$O ratio of 6.4$\pm$0.3 was already derived in Sect.\,4.1. This is the best known
isotope ratio in the nuclear region of NGC\,4945. Its value is higher than that measured in the
interstellar medium of the Milky Way or in the Sun ($^{18}$O/$^{17}$O=3.5 and 5.5, respectively;
see Penzias 1981) and can be explained by the peculiar isotopic composition of the ejecta from
massive stars that play an important role in a starburst environment.

For the central region of NGC\,4945, Chin et al. (1999) determined the $^{14}$N/$^{15}$N isotope
ratio on the basis of three HCN $J$=1--0 lines, those of HCN, H$^{13}$CN and HC$^{15}$N. Their measured
H$^{13}$CN/HC$^{15}$N line ratio of $\sim$2 combined with a $^{12}$C/$^{13}$C ratio of $\sim$50
resulted in $^{14}$N/$^{15}$N$\sim$100, a value that is much smaller than any value measured in the
galactic interstellar medium. As a consequence, massive stars must be a source of $^{15}$N. With
an H$^{13}$CN/HC$^{15}$N line temperature ratio of 2.1$\pm$0.3 (see Table~2) and a $^{12}$C/$^{13}$C
ratio of 50$\pm$10 we obtain $^{14}$N/$^{15}$N=105$\pm$26, in good agreement with Chin et al. (1999).

For the disk of the Galaxy (galactocentric distances 3\,kpc$\leq\,D_{\rm GC}\leq$9\,kpc) Chin et al.
(1996) found $^{32}$S/$^{34}$S ratios of (3.3$\pm$0.5)($D_{\rm GC}$/kpc)\,+\,4.1$\pm$3.1. The $^{32}$S/$^{34}$S
ratio in NGC\,4945 can be directly obtained from the CS/C$^{34}$S $J$=3--2 line intensity ratio, since
CS/$^{13}$CS may be close to the $^{12}$C/$^{13}$C isotope ratio so that CS line saturation should not
be very significant. In the `worst' case, i.e. if the CS/$^{13}$CS intensity ratio is 40 while
$^{12}$C/$^{13}$C$\sim$60 (Table~5 and Sect.\,6.1), the CS/C$^{34}$S line intensity ratio has to
be multiplied by 1.5 to obtain $^{32}$S/$^{34}$S. As a consequence we estimate $^{32}$S/$^{34}$S=13.5$\pm$2.5.
This corresponds to the value determined by Chin et al. (1996) in the innermost galactic disk ($D_{\rm GC}\sim$3\,kpc).
As a result we may have found not only in the inner galactic disk (Chin et al. 1996) but also in a nuclear
starburst environment a $^{32}$S/$^{34}$S ratio that is substantially smaller than that in the solar system.
So far this is the only nuclear starburst that was studied in $^{32}$S/$^{34}$S and the upper limit to the
$^{12}$C/$^{13}$C ratio is not strongly constrained (Sect.\,6.1). Therefore a clear statement on $^{34}$S
overproduction, compared to $^{32}$S, by massive stars cannot yet be given and more observational data
are needed to establish a clear picture. A summary of isotope ratios is given in Table~6.

\begin{table}
\begin{threeparttable}
\begin{scriptsize}
\begin{minipage}{180mm}
\caption[]{Isotope ratios\tnote{a)}}
\begin{flushleft}
\begin{tabular}{rr @{$\pm$} lr @{$\pm$} lllr @{$\pm$} l c}
\hline
 &\multicolumn{2}{c}{NGC\tnote{b)}} &
 \multicolumn{2}{c}{NGC\tnote{b)}}  &
 \multicolumn{1}{l}{M82\tnote{b)}}  &
 \multicolumn{1}{l}{Gal.\tnote{c)}} &
 \multicolumn{2}{c}{Solar\tnote{d)}} &
 \multicolumn{1}{c}{Solar}\\

 &\multicolumn{2}{c}{4945}          &
 \multicolumn{2}{c}{253}            &
                                    &
 \multicolumn{1}{l}{Center}         &
 \multicolumn{2}{c}{Circle}         &
 \multicolumn{1}{l}{Syst.}          \\
\hline \hline
$^{12}$C/$^{13}$C  & 50   &10  & 40  &10               &$>$40  &$\sim$25   &77  & 7  & 89      \\
$^{16}$O/$^{18}$O  &195   &45  &200  &50               &$>$90  &$\sim$250  &560 &25  &490      \\
$^{18}$O/$^{17}$O  &  6.4 & 0.3&  6.5& 1.0             &$\ga$8 &$\sim$3.5  &3.6 & 0.2&  5.5    \\
$^{14}$N/$^{15}$N  &105   &25  &\multicolumn{2}{c}{...}&$>$100 &$>$600     &450 &22  &270      \\
$^{32}$S/$^{34}$S  &13.5  & 2.5&\multicolumn{2}{c}{...}&...    &...        &32  & 7  & 22      \\
\hline
\end{tabular}
\begin{tablenotes}
\item[a)] Data taken from Penzias (1981), Henkel \& Mauersberger (1993), Henkel et al. (1993, 1998),
Wilson \& Rood (1994), Chin et al. (1996), Harrison et al. (1999) and this paper. For the Large
Magellanic Cloud (LMC), see Table~1 of Chin (1999).
\item[b)] Nuclear starburst environment
\item[c)] Isotope ratios in the central kpc of the Galaxy
\item[d)] Isotope ratios in the local interstellar medium
\end{tablenotes}
\end{flushleft}
\end{minipage}
\end{scriptsize}
\end{threeparttable}
\end{table}

\begin{table}
\begin{threeparttable}
\begin{scriptsize}
\begin{minipage}{180mm}
\caption[]{Column densities and densities\tnote{a)}}
\begin{flushleft}
\begin{tabular}{l r@{$\times$10}l r@{$\times$10}l}
\hline
Molecule& \multicolumn{2}{c}{$N$}          & \multicolumn{2}{c}{$n_{\rm H_2}$} \\
        & \multicolumn{2}{c}{(cm$^{-2}$)}  & \multicolumn{2}{c}{(cm$^{-3}$)} \\
\hline\hline
CO \tnote{b)}            &4.6      &$^{18}$  &4.0       &$^3$\\
$^{13}$CO                &9.2      &$^{16}$  &4.9       &$^3$\\
C$^{18}$O                &2.4      &$^{16}$  &2.6       &$^3$\\
C$^{17}$O                &3.7      &$^{15}$  &4.0       &$^3$\\
CN \tnote{c)}            &2.5      &$^{14}$  &$\sim$1.0 &$^{4}$\\
CS \tnote{d)}            &3.7      &$^{14}$  &5.0       &$^4$\\
C$^{34}$S                &2.7      &$^{13}$  &5.5       &$^4$\\
$^{13}$CS                &$\la$7.4 &$^{12}$  &5.5       &$^4$\\
SO \tnote{e)}            &1.0      &$^{14}$  &1.0       &$^5$\\
C$_2$H \tnote{f)}        &6.8      &$^{15}$  &\multicolumn{2}{c}{...}  \\
HCN \tnote{b)}           &1.3      &$^{15}$  &1.5       &$^5$\\
H$^{13}$CN               &2.6      &$^{13}$  &1.5       &$^5$\\
HC$^{15}$N               &9.3      &$^{12}$  &1.5       &$^5$\\
HCO$^+$ $^{\rm e,g)}$    &$\ga$1.9 &$^{13}$  &\multicolumn{2}{c}{...}  \\
HNC $^{\rm b,h)}$        &1.8      &$^{15}$  &\multicolumn{2}{c}{...}  \\
HN$^{13}$C \tnote{h)}    &3.5      &$^{13}$  &\multicolumn{2}{c}{...}  \\
N$_2$H$^+$ $^{\rm e,h)}$ &4.2      &$^{12}$  &\multicolumn{2}{c}{...}  \\
OCS \tnote{e)}           &$\la$5.2 &$^{14}$  &\multicolumn{2}{c}{...}  \\
ortho-H$_2$CO \tnote{e)} &1.0      &$^{14}$  &4.0       &$^5$\\
HNCO \tnote{e)}          &2.3      &$^{14}$  &1.6       &$^4$\\
C$_3$H$_2$ \tnote{e)}    &3.3      &$^{14}$  &\multicolumn{2}{c}{...}  \\
HC$_3$N \tnote{e)}       &7.0      &$^{13}$  &1.0       &$^5$\\
CH$_3$OH \tnote{e)}      &5.5      &$^{14}$  &1.1       &$^4$\\
CH$_3$CCH \tnote{i)}     &2.4      &$^{14}$  &\multicolumn{2}{c}{...}  \\
\hline
\end{tabular}
\begin{tablenotes}
\item[a)] For $T_{\rm kin}$=100\,K (see also Sect.~8); for assumed source sizes, see Sect.~4.
\item[b)] Assumed: $^{12}$C/$^{13}$C=50 (Sect.\,6.1)
\item[c)] Line saturation estimated by the relative strength of the fine structure components
\item[d)] $^{32}$S/$^{34}$S=13.5 (Sect.\,6.2)
\item[e)] Assumed: Optically thin emission (questionable at least in the case of HCO$^{+}$);
          for H$_2$CO, C01 estimated an ortho-para ratio $\sim$2.3.
\item[f)] Assumed: $T_{\rm ex}$=20\,K
\item[g)] Assumed: $T_{\rm ex}$=5\,K (see Sect.\,5.3)
\item[h)] $T_{\rm ex}$=8\,K, as for HCN
\item[i)] See Sect.\,5.6
\end{tablenotes}
\end{flushleft}
\end{minipage}
\end{scriptsize}
\end{threeparttable}
\end{table}

\section{Column densities and densities}

In Sect.\,4 column densities for several molecular species were determined. Making use of the isotope
ratios determined in Sect.\,6 and adding several molecular species, where measurements did not yet
justify an LVG treatment (Sect.\,5) provides a total of 24 molecular column densities adopting
$T_{\rm kin}$=100\,K and source sizes as outlined in Sect.\,4. The column densities do not change
significantly for $T_{\rm kin}$=50\,K. Changes in source size would, however, have a major impact with
column densities increasing with decreasing extent of the molecular emission. The assumption of similar
source sizes for all molecular species, close to those determined for CO and the 1.3~mm continuum
(Sect.~2.2), is most reasonable in view of missing interferometric data. A summary of estimated column
densities and densities is given in Table~7.

While the column densities of $^{13}$CS and OCS are based on tentative detections and are therefore upper
limits, the column densities of several species are lower limits in the case that line saturation effects
play a role. While this is unlikely in the case of HC$_3$N (Sect.\,4.3) and SO (Sect.\,4.4), this is less
clear for N$_2$H$^+$, HCO$^+$, H$_2$CO, HNCO, CH$_3$OH, C$_3$H$_2$ and CH$_3$CCH. Detections of rare
isotopomers would be needed to clarify the situation.

Comparing our column densities with those of C01 we note that some of the values in Table~7 are significantly
smaller than those quoted by C01. While in the case of CO and HCO$^+$ differences are almost negligible,
for CS, HCN and H$^{13}$CN the values differ by about one order of magnitude. We do not know the cause of
this effect but note that excitation temperatures determined by us for HCN are small and that our CS and
HCN data are more complete than those of C01.

H$_2$ densities are higher than expected. While densities of several 10$^3$\,cm$^{-2}$ deduced from CO
and of $\sim$10$^4$\,cm$^{-3}$ from CN and CH$_3$OH agree with estimates given by H94 and C01, most molecular
species indicate $n_{\rm H_2}\sim$10$^5$\,cm$^{-3}$. Densities for $T_{\rm kin}$=50\,K instead of 100\,K would
be even larger.  We thus conclude that the nuclear region of NGC\,4945 contains a prominent high density
component that is most convincingly supported by our multiline analysis of HC$_3$N (Sect.\,4.3). While the
uncertainty in $T_{\rm kin}$ and the choice of spherical or plan-parallel cloud geometry may introduce
deviations by factors of up to three from the values given in Table~7, we nevertheless find that most
densities are significantly larger than those estimated by C01.

\section{Relative abundances}

\subsection{$N(\rm H_2)$, M$(\rm H_2)$ and $X(\rm CO)$}

Since the column density of CO is relatively well determined for the inner $\sim$20$''$ of NGC\,4945,
an H$_2$ density can also be calculated adopting a [CO]/[H$_2$] abundance ratio of $\sim8\times10^{-5}$
(see Bradford et al. 2003). The total column density is then $N(\rm H_2)\sim$~6$\times$10$^{22}$\,cm$^{-2}$

To check the reliability of the H$_2$ column density, we use our
measurements of the 1.3~mm continuum. This continuum, described in
Sect.\,2.2, appears to arise from a similar volume than the
molecular gas and is thus interpreted as emission from dust grains
associated with molecular clouds in the nuclear environment. For
an estimate of $N$(H$_2$) we have to account for the contamination
of the measured `continuum' by the CO $J$=2--1 emission line (all
other lines contribute much less). With a sensitivity of 25\,Jy/K,
the 920\,K\,km\,s$^{-1}$ correspond to a flux of 17600\,Jy\,MHz.
Taking the width of the bolometer filter (30\,GHz), the
contamination of the measured continuum by CO $J=2-1$ in the
central beam is 0.60\,Jy or 43\% of the measured flux. This is
larger than the 14\% determined for the center of M\,82 (Thuma et
al. 2000). Assuming that a contamination of 43\% is representative
for the central emission in NGC\,4945, the pure continuum flux in
the central beam is  1.3\,Jy and 1.6\,Jy in the mapped region.

Following Mezger (1990), the H$_2$ column density can be obtained from
$$
N({\rm H_2})=0.97\times10^{15}cm^{-2}\frac{(S_{\nu}/{\rm Jy})\lambda_{\mu\rm m}^4}
{{({\theta_{\rm b}}/{\rm arcsec})^2}(Z/{\rm Z_\odot)}bT_{\rm d}}
\frac{{\rm e}^x-1}{x}.
$$
Here, $\theta_{\rm b}$ is the beam size, $S_\nu$ denotes the flux
density integrated over the beam, $Z$ represents the metallicity,
$T_{\rm d}$ is the dust temperature, and $x=14.4/\lambda_{\rm
mm}T_{\rm d}$. For the nuclei of external galaxies, $b$=1.9 (this
dimensionless parameter allows for changes in the dust absorption
cross section in various environments; see e.g. Mezger et al.
1987) may be justified, and $T_{\rm d}$ should be of order 50\,K
(IRAS 1989). If we assume that the metallicity is twice the solar
value (for galactic disk abundance gradients, see e.g. Hou et al.
2000), our pure continuum flux density in the central beam
(1.4\,Jy$\times$0.57$\sim$0.8\,Jy) translates into $N(\rm
H_2)=4.5\times10^{22}\,\rm cm^{-2}$, in good agreement with the
determination from CO. The main uncertainty, which we estimate to
be a factor of two in both directions, are the metallicity, the
factor $b$, and only to a lesser degree the dust temperature.

If we assume a column density of $N$(H$_2$) $\sim$
6$\times$10$^{22}$\,cm$^{-2}$ (this also compares well with X-ray
data suggesting $N$(H) $\sim$ 10$^{23}$\,cm$^{-2}$ for the nuclear
environment, while $N$(H) $\sim$ 4$\times$10$^{24}$\,cm$^{-2}$ is
found for the direct line-of-sight to the nucleus (Done et al.
2003)), $X$(CO)=$N$(H$_2$)/$\int T_{\rm CO 1-0}\,{\rm
d}v\,\eta_{bf}^{-1}$ (see Table~4) becomes
$\sim$3.5$\times$10$^{19}$\,cm$^{-2}$~[K~km$^{-1}$]$^{-1}$ (cf.
Mauersberger et al. 1996a).  This is a factor of $\sim$6.5 smaller
than the standard galactic $X$(CO) factor of
2.3$\times$10$^{20}$\,cm$^{-2}$~[K~km$^{-1}$]$^{-1}$. Such a low
conversion factor is consistent with values found in the central
region of other starburst galaxies (e.g. Mauersberger et al.
1996a, b; Solomon et al. 1997; Wei{\ss} et al. 2001b).

Table~8 gives fractional abundances of molecular species with respect to $N(\rm H_2)$ studied in
the starburst galaxies NGC\,4945, NGC\,253 and M\,82 and toward prototypical galactic molecular clouds,
the Orion Hot Core, the Orion Ridge and TMC-1.

\subsection{NGC\,4945, Orion and TMC-1}

Agreement between the fractional abundances derived from the
central region of NGC\,4945 and the galactic sources is obtained
for CS, N$_2$H$^+$, H$_2$CO, HC$_3$N, and CH$_3$CCH. A multilevel
HC$_3$N study of the Sgr\,A molecular clouds near the center of
the Milky Way provides, on smaller linear scales than for
NGC\,4945, similar excitation conditions, i.e. $n_{\rm H_2}$
$\sim$ 10$^5$\,cm$^{-3}$ for $T_{\rm kin}$=80\,K (Walmsley et al.
1986). For SO and HCN, the abundances agree with those for the
Orion Ridge and TMC-1, but are lower than the enhanced abundances
of the Orion Plateau and Hot Core, respectively. The C$_2$H abundance
seems to be much higher in NGC\,4945 than in the galactic sources,
but excitation and optical depths are poorly constrained. The HNCO
abundance in NGC\,4945 is consistent with those of galactic sources
containing warm ($T_{\rm kin}$ $>$ 10\,K) gas, while the abundance
of cyclic C$_3$H$_2$ seems to be similar to that obtained in the TMC-1.
The CH$_3$OH abundance is higher than in cold quiescent galactic clouds,
but is lower than in the Orion compact ridge.

With respect to warm galactic sources, HNC is significantly
enhanced. In the galactic disk, interstellar $N$(HNC)/$N$(HCN)
abundance ratios cover a range of at least two orders of
magnitude. In quiescent dark clouds the abundance ratio is close
to or larger than unity (e.g. Churchwell et al. 1984; Harju 1989),
while in warmer and denser clouds associated with massive star
formation the ratio can decrease to values of $\sim$0.01 (e.g.
Goldsmith et al. 1981, 1986; Schilke et al. 1992). It is thus
highly surprising to find $N$(HNC)/$N$(HCN)$\sim$1 in the warm
environment of a nuclear starburst, like that in NGC\,4945.

\begin{table}
\begin{threeparttable}
\begin{scriptsize}
\begin{minipage}{180mm}
\caption[]{Logarithmic fractional abundances\tnote{a,b)}}
\begin{flushleft}
\begin{tabular}{l r rrrrr}
\hline
Molecule & \multicolumn{1}{c}{NGC} & \multicolumn{1}{c}{NGC}& M\,82& \multicolumn{1}{c}{Orion}&
          \multicolumn{1}{c}{Orion}& \multicolumn{1}{c}{TMC-1} \\
         & \multicolumn{1}{c}{4945}& \multicolumn{1}{c}{253}&      & \multicolumn{1}{c}{Hot}  &
           \multicolumn{1}{c}{Ridge}&                           \\
         & \multicolumn{1}{c}{         }&         &      &           \multicolumn{1}{c}{Core}&            &      \\
\hline\hline
CO\tnote{c)}      &-4.1        &-4.1        &-4.1     &-3.9      &-4.3            &-4.2   \\
$^{13}$CO         &-5.8        &-5.7        &$\la$-5.7&...       &...             &...    \\
C$^{18}$O         &-6.4        &-6.4        &$\la$-6.1&...       &...             &...    \\
C$^{17}$O         &-7.2        &-7.2        &$\la$-7.0&...       &...             &...    \\
CN                &-8.4        &-8.6        &-8.4     &$\la$-9.5 &-8.5            &-9.2   \\
CS                &-8.2        &-8.7        &-7.9     &-8.2      &-8.6            &-8.6   \\
C$^{34}$S         &-9.4        &...         &...      &...       &...             &...    \\
$^{13}$CS         &$\la$-9.9   &-10.3       &...      &...       &...             &...    \\
SO                &-9.0        &...         &...      &$\la$-7.7 &$\la$-9.0       &-9.0   \\
C$_2$H            &-6.9        &...         &-6.8     &$\la$-9.5 &-8.3            &-8.6   \\
HCN               &-7.7        &-8.0        &-8.0     &-6.5      &-8.3            &-8.4   \\
H$^{13}$CN        &-9.4        &-9.6        &...      &...       &...             &...    \\
HC$^{15}$N        &-9.8        &...         &...      &...       &...             &...    \\
HCO$^+$           &$\ga$-9.5   &-7.7        &-8.3     &-8.6      &-8.6            &-8.1   \\
HNC               &-7.5        &-8.3        &-8.7     &-9.0      &-9.3            &-7.9   \\
HN$^{13}$C        &-9.3        &-9.9        &...      &...       &...             &...    \\
N$_2$H$^+$        &-10.2       &-9.9        &-9.8     &...       &-10.2           &-10.0  \\
OCS               &$\la$-8.1   &-7.5        &$<$-7.9  &...       &-8.5\tnote{d)}  &...    \\
H$_2$CO\tnote{e)} &-8.7        &-7.9        &-7.5     &-7.6      &-7.8\tnote{d)}  &-7.9   \\
HNCO              &-8.6        &-8.4        &-8.1     &-8.2      &$\la$-8.7       &-10.1  \\
C$_3$H$_2$        &-8.3        &-9.2        &-8.7     &...       &-10.1           &-8.0   \\
HC$_3$N           &-9.0        &-7.9        &-8.0     &-8.8      &-9.9            &-8.9   \\
CH$_3$OH          &-8.1        &-7.9        &$<$-8.7  &-6.5      &-6.9\tnote{d)}  &-9.1   \\
CH$_3$CCH         &-8.4        &-7.8        &-7.3     &...       &-8.5            &-8.7   \\
\hline
\end{tabular}
\begin{tablenotes}
\item[a)] Most fractional abundances  of the TMC-1 were taken from Pratap et al. (1997, their weighted
mean; see also Leung et al. 1984; Millar \& Freeman 1984; Madden et al. 1989). Orion data were taken
from Blake et al. (1987) and Comito (2003). Fractional abundances of NGC\,253 and M\,82 ($N$(H$_2$)
= 4.0$\times$10$^{22}$\,cm$^{-2}$ and 1.0$\times$ 10$^{22}$\,cm$^{-1}$, respectively; Mauersberger et al.
2003) are, whenever possible, based on spectra from the 30-m telescope on Pico Veleta/Spain (see Henkel
et al. 1988, 1993, 1998; Mauersberger \& Henkel 1989; Mauersberger et al. 1990, 1991, 1995; Nguyen-Q-Rieu et
al. 1991, 1992; H{\"u}ttemeister et al. 1995, 1997; Harrison et al. 1999; Oike et al. 2004), complemented
by higher excited lines measured with similar beam size at the HHT (Heinrich Hertz Telescope) on Mt. Graham/USA
(Mauersberger et al. 2003). For NGC\,4945, see this paper ($N$(H$_2$) = 6$\times$10$^{22}$\,cm$^{-2}$);
for N$_2$H$^+$, see Sage \& Ziurys (1995).
\item[b)] In most cases excitation conditions were assumed to be the same in NGC\,4945, NGC\,253 and M\,82.
This is a simplification in view of the discussion about HC$_3$N in Sect.\,8.3. The CN opacity correction
derived in Sect.\,5.1.1 for NGC\,4945, +0.2\,dex, was however not applied to NGC\,253 and M\,82 (for the
CN data, see Henkel et al. 1988). While HCO$^{+}$ emission was assumed to be optically thin in
NGC\,4945 (Sect.\,5.3), a small HCO$^+$/H$^{13}$CO$^{+}$ line ratio ($\sim$10) and $^{12}$C/$^{13}$C
$\sim$ 40 (Henkel et al. 1993) yield $\tau_{\rm HCO^+}$ $\sim$ 4 for NGC\,253. Thus for NGC\,253
0.6\,dex were added to the column density calculated assuming optically thin emission. Fractional
N$_2$H$^+$ abundances were obtained from 12-m Kitt Peak data (Sage \& Ziurys 1995) with a large beam
size, that reduces their beam averaged column density significantly. This is compensated by the high
rotational temperature ($T_{\rm rot}$ = 30\,K versus 8\,K in Table~7). $N$(N$_2$H$^+$) in NGC\,253
and M\,82 and $N$(HCN), $N$(HNC) and $N$(C$_2$H) in M\,82 may suffer from optical depth and excitation
effects that cannot be quantified. For details concerning column densities in NGC\,4945, see Table~7.
From the OCS column densities, derived by Mauersberger et al. (1985) for 11$''$ sized regions, 0.6\,dex
were subtracted to adjust $N$(OCS) to a source size of 20$''$. While there exists an HC$_3$N multilevel
study for NGC\,253 (Mauersberger et al. 1990) similar to the one outlined in Sect.\,4.3 for NGC\,4945,
only the HC$_3$N $J$-10--9 line was reported from M\,82 (Henkel et al. 1988). Adopting the excitation
conditions of NGC\,4945 yields a logarithmic fractional abundance of --8.7, while excitation conditions
similar to NGC\,253 yield --7.3 for M\,82. The mean of these two values is given.
\item[c)] For the extragalactic sources, the fractional CO abundance was taken from Frerking et al.
(1982). The corresponding abundances for the galactic sources are based on direct measurements (e.g.
Blake et al. 1987).
\item[d)] Orion Compact Ridge
\item[e)] Galactic abundances include para- and ortho-H$_2$CO. For the extragalactic sources
ortho-H$_2$CO column densities (Table~7; H{\"u}ttemeister et al. 1997) were upgraded by 0.1\,dex.
\end{tablenotes}
\end{flushleft}
\end{minipage}
\end{scriptsize}
\end{threeparttable}
\end{table}

\subsection{NGC\,4945, NGC\,253 and M\,82}

NGC\,253, M\,82~(NGC\,3034) and NGC\,4945 are with $D$=2.5--4.0\,Mpc the three nearest starburst
galaxies. All of them show a rich molecular spectrum and infrared luminosities of a few times
10$^{10}$\,L$_\odot$. Line intensities in NGC~253 and NGC\,4945 can be similar while those towards
M~82 tend to be slightly lower (compare e.g. our Table~2 with Table~1 in Sage \& Ziurys 1995). While
the starburst in NGC\,253 is compact, mostly originating from a 60\,pc sized region slightly southwest
of the nucleus (Telesco et al. 1993), the starburst in M\,82 is more extended exhibiting two lobes
at a distance of $\sim$150--250\,pc from the dynamical center that may be part of a molecular ring
(Mao et al. 2000).

Not only with respect to size and overall morphology but also with
respect to molecular abundances, the nuclear regions of NGC\,253
and M\,82 are quite distinct. While abundances of CN, CS, HCN,
HNC, C$_3$H$_2$ and CH$_3$CCH are similar, SiO, HNCO, CH$_3$OH and
CH$_3$CN are clearly underabundant in M\,82 (see Mauersberger et
al. 1991, 2003; Nguyen-Q-Rieu et al. 1991, 1992; Henkel et al.
1993; Mauersberger \& Henkel 1993; H\"uttemeister et al. 1997;
Garc\'{\i}a-Burillo et al. 2001). Recently, an underabundance in
NH$_3$ was also demonstrated (Wei{\ss} et al. 2001a; Mauersberger
et al. 2003), while HCO was detected in M\,82, not in NGC\,253
(Garc\'{\i}a-Burillo et al. 2002). According to Takano et al.
(1995), molecules formed by ion-molecule and neutral-neutral
reactions with a low activation energy barrier are seen in both
galaxies, while species requiring neutral-neutral reactions with
significant activation energy or dust grain related reactions are
abundant only in NGC\,253.

In order to explain the chemical discrepancies, Mao et al. (2000)
proposed the presence of a warm diffuse interstellar medium that
is responsible for the bulk of the CO emission in M\,82. This
medium is a consequence of cloud-cloud collisions, shocks, high
gas pressure, high stellar densities and numerous young massive
stars that may have led to the evaporation of vast amounts of
dense cool gas creating a highly fragmented molecular cloud
debris. While the use of their LVG code does not necessarily lead
to an area filling factor that is similar to the volume filling
factor and while not all of their PDR model input parameters are
self-consistent, more recently published data convincingly confirm
their proposed scenario. CO is highly excited and must arise
predominantly from gas with kinetic temperatures well in excess of
50\,K (Mao et al. 2000). The rotation temperature of NH$_3$ is,
however, only $\sim$30\,K (Wei{\ss} et al. 2001a; Mauersberger et
al. 2003), implying $T_{\rm kin}$$\sim$50\,K, and HCN data suggest
temperatures of order $T_{\rm kin}\sim$20--60\,K (Seaquist et al.
1998). Wei{\ss} et al. (2001a) also find that the fractional
NH$_3$ abundance decreases from low values in the lobes to even
smaller ones near the dynamical center.

Overall 60\,$\mu$m/100\,$\mu$m dust temperatures for NGC\,253 and M\,82 are (as the total infrared
luminosities) quite similar. They are of order 50\,K for a modified Planck function $S_{\nu}\ =
\epsilon_{\nu}B_{\nu}(T_{\rm d})$, where $T_{\rm d}$ is the dust temperature, $B_{\nu}(T_{\rm d})$
the Planck function and $\epsilon_{\nu}$ the emissivity of the dust grain population at frequncy $\nu$.
$\epsilon_{\nu}\propto\nu^\beta$, $\beta$=1. In M\,82 it thus appears that NH$_3$, HCN and other
high density tracers arise from dense gas with kinetic temperatures close to the dust temperature,
that is too low for the synthesis of significant amounts of SiO and CH$_3$CN or for NH$_3$ and
CH$_3$OH evaporation from dust grains. The lower density tracer CO, however, predominantly originates
from the warm UV irradiated surfaces of molecular clouds. This view is further supported by the
comparatively larger amount of molecular gas at $T_{\rm kin}\sim$150\,K, about ten times as much as
in NGC\,253 (Rigopoulou et al. 2002). At these temperatures, NH$_3$ is evaporated from dust grains.
Since we see NH$_3$ at such temperatures in NGC\,253 but not in M\,82, the warm ($T_{\rm kin}\sim$150\,K)
molecular gas in M\,82 must be tenuous, i.e. column densities of individual molecular filaments are not
high enough to shield the molecules from rapid photodissociation. Since HCO is observed in the Milky Way
at the interfaces between molecular and ionized gas, the non-detection of HCO in NGC\,253 and detailed maps
of HCO in M\,82 (Garc\'{\i}a-Burillo et al. 2002) further strengthen the idea that the molecular region of
M\,82 can be viewed as a giant PDR. Apparently, the starburst in M\,82 represents a later stage of evolution
than the starburst in NGC\,253.

The crucial question then is whether the nuclear region of NGC\,4945 resembles more closely that of NGC\,253,
that of M\,82 or whether NGC\,4945 represents the first known example of a starburst in yet another evolutionary
stage. Based on the weakness of B$\gamma$ emission Koornneef (1993) coined the term `Postburst Infrared Galaxy'
for objects like NGC\,4945. Therefore similar physical and chemical conditions might be expected in the nuclear
environment of M\,82 and NGC\,4945. Ott et al. (2001), however, identified a bar-like feature in NGC\,4945 and
claimed that large amounts of molecular material can still reach the nuclear region providing sufficient fuel
for an ongoing starburst.

Table~8 provides some clues to the stage of the starburst in
NGC\,4945. Apparently, fractional abundances of CH$_3$OH and HNCO
are close to those of NGC\,253, i.e. NGC\,4945 does ${\it not}$
show the underabundances that characterize the molecular spectrum
of M\,82. For NGC\,4945 and NGC\,253, carbon and oxygen isotope
ratios appear to be identical within the error limits, while the
ratios are less well constrained for M\,82 (Table~6). An important
test, the measurement of thermal SiO, HCO and CH$_3$CN, is,
however, still required to obtain a complete picture. While we may
have seen some traces of SiO in NGC\,4945 (see Fig.\,5 and
Sect.\,5.3), better data are needed for a comparison.

An evaluation of molecular line excitation can be made for
HC$_3$N, comparing our data with the HC$_3$N multilevel study of
Mauersberger et al. (1990) towards NGC\,253. In the latter galaxy,
excitation is significantly higher than in NGC\,4945, i.e. there
is a high excitation component leading to a detection of the
$J$=26--25 line. For NGC\,4945, our model calculations
(Sect.\,4.3) clearly indicate that the line is undetectable even
if a 30-m sized telescope had been used.

\subsection{CN, HNC and HCN in NGC\,4945 and other galaxies}

H{\"u}ttemeister et al. (1995) and Aalto et al. (2002) compared CN
$N$=1--0 and 2--1, HNC $J$=1--0 and HCN $J$=1--0 line intensities
in a representative sample of galaxies, including targets that
resemble the Galaxy (e.g. IC\,342), weak starburst galaxies (e.g.
NGC\,253 and M\,82), luminous infrared galaxies (LIRGs; e.g.
Arp\,299) and ultraluminous infrared galaxies (ULIRGs; e.g.
Arp\,220). As already mentioned in Sect.\,8.2, galactic HNC/HCN
abundance ratios vary by at least two orders of magnitude (Schilke
et al. 1992). This is, however, not fully seen in the extragalactic
line intensity ratios, (1) because particularly high HCN abundances
lead to optically thick $J$=1--0 lines with intensities that do not
fully reflect the column density and (2) because the surveyed regions
of $\la$1\,kpc in size smear out any extreme values on small spatial
scales. Thus [HNC]/[HCN] $\la$ $I$(HNC)/$I$(HCN) $\la$ 1 with [X] and
$I$(X) denoting the abundance (relative to H$_2$) and intensity of the
respective molecule. Equal excitation is assumed and is supported by
similar molecular constants and electric dipole moments.

Towards NGC\,4945, the HNC abundance is high.
$I$(HNC)/$I$(HCN)$\sim$0.5 is near the average of the values
determined by H{\"u}ttemeister et al. (1995) and Aalto et al.
(2002). The ratio of the optically thin lines of the rare
isotopomers HN$^{13}$C and H$^{13}$CN indicates, however, that
[HNC]$\sim$10$^{-7.5}$. This is exemplified by the relatively
small HNC/H$^{13}$NC line intensity ratio in NGC\,4945 (see Table
2) and the relatively large one in NGC\,253 (Henkel et al. 1993),
indicating lower optical depths in the latter case, since the
$^{12}$C/$^{13}$C carbon isotope ratios appear to be similar in
both galaxies.

\begin{table}
\begin{scriptsize}
\caption[]{Relevant line intensity ratios}
\begin{flushleft}
\begin{tabular}{r@{/}lc}
\hline
CN $N$=1--0         & HCN $J$=1--0          & 1.79 \\
CN $N$=2--1         & HCN $J$=1--0          & 1.31 \\
HCO$^+$ $J$=1--0    & HCN $J$=1--0          & 1.07 \\
HNC $J$=1--0        & HCN $J$=1--0          & 0.54 \\
HN$^{13}$C $N$=1--0 & H$^{13}$CN $J$=1--0   & 1.10 \\
\hline
\end{tabular}
\begin{tablenotes}
\item[a)] Taken from Table 2. Uncertainties amount to $\sim$$\pm$15\%.
\end{tablenotes}
\end{flushleft}
\end{scriptsize}
\end{table}

As already mentioned in Sect.\,8.2, high HNC/HCN line intensity
and abundance ratios (near unity) are a characteristic property of
dark clouds with small temperature ($T_{\rm kin}$$\sim$10\,K) and
moderate density ($n$(H$_2$)$\sim$10$^{4-5}$\,cm$^{-3}$). At
higher $T_{\rm kin}$ values and densities HNC may more efficiently
react e.g. via HNC+O$\rightarrow$CO+NH and HNC+H$\rightarrow$HCN+H
that shifts the chemical balance between HNC and HCN in favor of
the latter species. Recently, however, Herpin \& Cernicharo (2000)
showed that in the warm ($T_{\rm kin}$$\sim$250\,K) and dense
($n$(H$_2$)$\ga$10$^{7}$) circumstellar torus of the
protoplanetary nebula CRL\,618 HNC and HCN are similarly abundant.
Aalto et al. (2002) proposed that [HNC]/[HCN]$\sim$1 is not only
typical for dark clouds but may also characterize warm regions as
long as the ion abundance is high. Protonized HCN and HNC, i.e.
HCNH$^{+}$, will then recombine to produce HCN and HNC in equal
quantities.

A cool molecular halo of moderate density with particularly high HNC abundance that surrounds the nuclear region
was alternatively proposed by H{\"u}ttemeister et al. (1995). This is, however, not a viable explanation for
all sources, since our study of NGC\,4945 is confined to the inner $\sim$20$''$ and a cool moderately dense
foreground layer from the outer disk (NGC\,4945 is viewed almost edge-on, see Sect.\,1) would have been
seen in other molecular lines, most notably those of CO. Another alternative, a cool moderately dense gas
component within the nuclear environment does also not explain all observations. While such gas is common in
the central kiloparsec of our Galaxy (e.g. H{\"u}ttemeister et al. 1997), ULIRGs like Mrk\,231 do not exhibit
a cool dust component that would be associated with such a hypothesized cool source of HNC emission.

Aalto et al. (2002) outlined an evolutionary scenario starting
with bright HCN emission from dense quiescent warm gas with little
CN and HNC, reaching a PDR-phase with $I$(CN)$\ga$$I$(HCN) and ending
with bright HNC but little CN, when the starburst has evolved
beyond a strong radiative impact from young massive stars, while
the cloud's chemistry is still dominated by ion-neutral reactions
at moderate gas density and relatively high electron abundance.
NGC\,4945 is in the intermediate phase, with $I$(CN)$>$$I$(HCN)
(see Table 9), like NGC\,253 (see Henkel et al.  1993), but with
considerably more HNC, suggesting a later stage of evolution.
M\,82 is also characterized by $I$(CN)$>$$I$(HCN), while the
optical depth of HNC remains poorly constrained (Henkel et al.
1998). The starburst in M\,82 is in a later stage of evolution,
because here it is not apparent that large amounts of fresh
molecular fuel continue to reach the central region. The $I$(CN
$N$=1--0 $J$=3/2--1/2)/$I$(HCN $J$=1--0) ratio may serve as an
indicator for the importance of PDRs (e.g. Greaves \& Church 1996;
Rodriguez-Franco et al. 1998). In the galactic nuclear region, the
ratio is $\sim$0.3 (Paglione et al. 2002). For NGC\,253, NGC\,4945
and M\,82, we find 0.8, 1.1 and 1.3, respectively.

\section{Energetics}

The heating of gas in the nuclear environment of spiral galaxies
is one of the major puzzles in our understanding of the
interstellar medium. A variety of potential heating mechanisms has
been proposed but to discriminate between the various physical
processes has turned out to be extremely difficult. Proposed were
(1) cosmic ray heating (e.g. G\"usten et al. 1981; Bradford et al.
2003), (2) stellar ultraviolet heating via PDRs (e.g. Hollenbach
\& Tielens 1997; Mao et al. 2000; Schulz et al. 2001;
Garc\'{\i}a-Burillo et al. 2002), (3) heating by X-rays, e.g. from
the active galactic nucleus (AGN) resulting in an X-ray dominated
region (XDR; e.g. Lepp \& Dalgarno 1996; Usero et al. 2004), (4)
dynamical heating by shocks (e.g. Mart\'{\i}n-Pintado et al. 1997,
2001; H\"uttemeister et al. 1998; Garc\'{\i}a-Burillo et al. 2000,
2001) triggered by outflowing material from massive stars or
active galactic nuclei, by cloud-cloud collisions, by tidal forces
or by clouds gravitationally dominated by the mass distribution of
the embedded stars (e.g. Downes et al. 1993; Mauersberger et al.
1996b) as well as (5) by ambipolar diffusion or ion slip that
might be particularly significant in the potentially strong
magnetic fields of the nuclear environment of spiral galaxies
(e.g. Scalo 1977).

In local dark clouds, cosmic rays are a major source of energy,
yielding kinetic temperatures of 5--10~K alone. With a detailed
chemical model, also involving cooling by the fine structure lines
of O and C and by CO, Farquhar et al. (1994) find for cosmic ray
heating $T_{\rm kin}$=6.7~$S^{0.31}$~K, with the factor $S$
denoting the enhancement ($S>$1) of the cosmic ray flux over the
solar system value. Adopting $S$$\sim$10$^3$ for sources like
NGC\,253, M\,82 and NGC~\,4945 (e.g. Bradford et al. 2003; for
supernova rates, see Ulvestad \& Antonucci 1997), we thus obtain
$T_{\rm kin}$$\sim$50~K, in good agreement with the temperature of
the dust in the starburst environment of all three galaxies. This
is consistent with a scenario for M~82 that assumes a significant
contribution to the overall heating by cosmic rays for the dense
deeply embedded gas, while the warmer CO is heated by PDRs (see
Sect.\,8.2). In NGC\,253, however, both CO (Bradford et al. 2003)
and NH$_3$ (Mauersberger et al. 2003) arise from regions
characterized by $T_{\rm kin}\ga$100~K, so that at least for these
two molecules, cosmic ray heating appears not to be sufficient. So
far, NGC\,4945 is less thoroughly studied. Detections of NH$_3$
and of highly excited CO lines have not yet been obtained. To
confirm or reject our finding that NGC\,4945 appears to resemble
more NGC\,253 than M\,82, higher excited ($J>$3) CO lines and
NH$_3$ should also be measured.

Since high $J$ CO and NH$_3$ line data are not yet available,
indications for high molecular cloud temperatures are still rare
for NGC\,4945. Perhaps the strongest hint for the presence of warm
molecular gas at mm-wavelengths is our HC$_3$N multilevel study
(Sect.\,4.3), where the assumption of a `normal' density also seen
in most other molecular tracers, $n_{\rm
H_2}\la$10$^5$\,cm$^{-3}$, leads to a kinetic temperature of
$T_{\rm kin}\ga$100~K. While the presence of gas at such
temperatures is a typical property of PDRs, HC$_3$N is a tracer of
deeply embedded molecular material that may be less affected by UV
heating than CO. Furthermore, HCO, a tracer of PDRs, appears to be
weak in NGC\,4945 (Sect.\,5.3). While PDRs will certainly exist in
the nuclear environment of NGC\,4945, they appear not to be
widespread enough to cause heating on large spatial scales.
Heating by shocks or ambipolar diffusion are thus more likely the
physical processes that may heat the gas on large scales in the
nuclear starburst environment of NGC\,4945. In view of the
presence of shocks that are expected to be associated with the bar
identified by Ott et al. (2001), the measurement of a well
calibrated SiO $J$=3--2 ($v$=0) profile would be highly desirable.

An alternative heating source could be X-ray radiation, leading to
a high ionization rate (Lepp \& Dalgarno 1996; Usero et al. 2004).
Here we expect that the 20$''$ region, corresponding to a linear
scale of 390\,pc, is too large to be dominated by the AGN, which
is consistent with an $I$(HCN)/$I$(CO) line intensity ratio that
is far from the extreme value measured toward the central region
of NGC\,1068 (Sternberg et al. 1994). The observed abundances of
CO, HCO$^{+}$ and HCN match those calculated by Lepp \& Dalgarno
(1996) for an ionization rate per hydrogen atom, $\xi$/$n$(H), of
$\sim$10$^{-19...-18}$\,cm$^{-3}$\,s$^{-1}$, but the measured CN
abundance (see Table 8) is an order of magnitude too low.

\section{Conclusions and prospects for future research}

Toward the nucleus of the southern starburst galaxy NGC\,4945,
80 lines from a total of 19 molecules have been observed, including
49 detected line features, 9 tentatively detected transitions and 22
undetected lines.

Two velocity components, one at $\sim$450\,km\,s$^{-1}$, the other
at $\sim$710\,km\,s$^{-1}$, characterize many but not all spectra.
The former component is wider; the latter component is higher
excited as is indicated by HCN, HCO$^+$ and CN spectra. While no
maps are presented, CO/$^{13}$CO line intensity ratios as a
function of velocity are consistent with a ring-like morphology of
the emitting gas.

Using the LVG or LTE approximation, H$_2$ densities and fractional
abundances of 24 molecular species (including isotopomers) were
calculated. Many of these species indicate the presence of a
prominent high density interstellar gas component characterized by
$n_{\rm H_2}$$\sim$10$^5$\,cm$^{-3}$. Excitation temperatures are
in some cases quite low and reach 3--4\,K in the case of CN. HNCO
data indicate a mm-wave background radiation field of $\la$30\,K.

Based on column densities derived from CO and from the 1.3~mm dust
continuum (2.8$\pm$0.3\,Jy), the H$_2$ to
integrated CO $J$=1--0
line intensity conversion factor `$X$' becomes
$\sim$3.5$\times$10$^{19}$\,cm$^{-2}$~[K~km$^{-1}$]$^{-1}$.
This is a factor $\sim$7 smaller than values found for the
galactic disk.

Calculated fractional abundances are compared with abundances
observed toward the starburst galaxies NGC\,253 and M\,82 and
selected galactic sources. The chemical properties of the gas in
NGC\,4945 resemble more those in NGC\,253 than in M\,82. There
are, however, differences between NGC\,4945 and NGC\,253. One is
excitation. HC$_3$N is more highly excited in NGC\,253.
Furthermore there is an `overabundance' of HNC in the nuclear
environment of NGC\,4945. The HNC/HN$^{13}$C $J$=1--0 line
intensity ratio is with $\sim$11 particularly small indicating a
high optical depth ($\sim$5) in HNC $J$=1--0. Thus NGC\,4945 is
one of the few known starburst galaxies with an HNC/HCN abundance
ratio $\sim$1 that may be caused by a high abundance of ions. A
detailed analysis of CN, HCN and HNC line intensities and
abundances suggests that the starburst in NGC\,4945 is in a stage
of evolution that is intermediate between those of NGC\,253 and
M\,82.

Carbon, nitrogen, oxygen and sulfur isotope ratios are determined.
Carbon and oxygen isotope ratios in the nuclear environment of
NGC\,4945 and NGC\,253 appear to be almost identical, while ratios
from M\,82 are less well constrained. High $^{18}$O/$^{17}$O, low
$^{16}$O/$^{18}$O and $^{14}$N/$^{15}$N and perhaps also low
$^{32}$S/$^{34}$S ratios (6.4$\pm$0.3, 195$\pm$45, 105$\pm$25 and
13.5$\pm$2.5, respectively) are characteristic properties of a
starburst environment, where massive stars have had sufficient
time to affect the isotopic composition of their surrounding ISM.
The $^{12}$C/$^{13}$C ratio is about twice as large as in the
central region of the Milky Way, suggesting the presence of
inflowing gas.

Much work still remains to be done. Determining isotope ratios of
sulfur in the nuclear environment of starburst galaxies offers a
unique chance to study the isotopic composition of sulfur ejecta
from massive stars undergoing oxygen burning. Obtaining a more
complete view onto the $^{32}$S/$^{34}$S ratio in NGC\,4945 and
estimates of $^{32}$S/$^{34}$S in NGC\,253, M\,82 and perhaps
Arp\,220 would thus provide important insights into aspects of
stellar nucleosynthesis. Detecting $^{13}$CN and obtaining a
definite detection of $^{13}$CS would further constrain the
$^{12}$C/$^{13}$C and $^{16}$O/$^{18}$O ratios in the nuclear
region of NGC\,4945 and would thus also be fundamental for a
better understanding of the chemical and physical conditions in
a starburst environment.

In spite of the large number of detections, our line survey is not
complete. While it appears that the chemical composition of the
gas in the central part of NGC\,4945 is more resembling NGC\,253
than M\,82, three important tracers have not yet been convincingly
measured: thermal SiO, CH$_3$CN and HCO. Furthermore, NO could be
an interesting tracer for X-ray related chemistry. If our finding
of similar chemical composition is valid for NGC\,4945 and
NGC\,253, SiO and CH$_3$CN should be detectable in NGC\,4945,
while HCO should be weak.

Our $^{13}$CO $J$=3-2 line profile is, with respect to intensity and
lineshape, incompatible with the other CO transitions observed.
This also holds for the lineshape of the C$_3$H$_2$ 2$_{12}$--1$_{01}$
transition with respect to the other C$_3$H$_2$ line profiles. Another
measurement with careful pointing is thus needed to obtain
better agreement or to confirm the inconsistencies.

By far the most important lack of observational data refers,
however, to the absence of interferometric maps in NGC\,4945.
Molecular measurements are so far constrained to linear scales of
several 100\,pc. It is still not known why some species show an
almost Gaussian line profile (e.g. H$_2$CO), while others show two
pronounced peaks at $\sim$450 and 710\,km\,s$^{-1}$ (e.g. CN).
High angular resolution maps could determine the location,
kinematics and spatial distribution of the distinct velocity
components. Where is the region with HNC/HCN$>$1 line intensity
ratios located? Will high resolution HNC and HCN line maps provide
important constraints for chemical models of the starburst
environment and the role of the AGN? And what is the distribution
of the HNC/HN$^{13}$C line intensity ratio that allows us to
estimate optical depths? Can SiO, HCO and H$^{13}$CO$^+$ be
disentangled from each other as it was possible in the case of
M\,82? In the near future, interferometric maps of ATCA
(Australian Telescope Compact Array) and SMA (SubMillimeter Array)
will be available and single-dish telescopes like APEX (Atacama
Pathfinder Experiment) will provide data on highly excited
spectral lines that are crucial to identify the major heating
source(s). The goal of this paper is to provide a reliable
observational basis for these measurements.

\begin{acknowledgements}
We wish to thank C.~C.~Chiong and C.~M.~Walmsley for providing an
LVG code for SO and S.~Leurini for the use of her CH$_3$OH LVG
code. We also acknowledge the use of HNCO collision rates provided
by S.~Green (http://www.giss.nasa.gov/data/mcrates/\#hnco) and of
CH$_3$OH collision rates provided by D.~Flower. We acknowledge
useful discussions with S.~Aalto, E. Hails and K.~Menten and thank
S. Garc\'{\i}a-Burillo for critically reading the manuscript.
M.~W. acknowledges support by the exchange program between the
Chinese Academy of Sciences and the Max-Plank-Gesellschaft, and
partly by grants 10133020 from NSFC \& G19990754 from CMST.
J.~B.~W. and M.~H.~C. acknowledge the financial support supplied
by the Australian Government's Access to Major Research Facilities
Program (AMRFP) for travel to the SEST.
\end{acknowledgements}

\end{document}